\documentclass[sigconf]{acmart}

\usepackage{balance}
\usepackage{tabulary}
\usepackage{multirow}
\usepackage{booktabs} 
\usepackage{array}
\usepackage[]{algorithm2e}

\makeatletter
\g@addto@macro{\UrlBreaks}{\UrlOrds}
\makeatother

\setcopyright{none}

\acmDOI{10.475/123_4}

\acmISBN{123-4567-24-567/08/06}

\newcolumntype{?}{!{\vrule width 1pt}}

\newcommand{\confirmedbugnum}{thirty-nine}
\newcommand{\confirmednpd}{seventeen}
\newcommand{\confirmeduaf}{ten}
\newcommand{\confirmedrl}{eleven}
\newcommand{\confirmedsae}{one}

\newcommand{\toolname}{\textsf{Catapult}}
\newcommand{\pin}{\textsf{Pinpoint}}
\newcommand{\fortify}{\textsf{Fortify}}
\newcommand{\saturn}{\textsf{Saturn}}
\newcommand{\calysto}{\textsf{Calysto}}

\newcommand{\fbinfer}{\textsf{Infer}}
\newcommand{\clang}{\textsf{CSA}}
\newcommand{\tricoder}{\textsf{Tricoder}}

\newcommand{\pc}{\textsf{pc}}
\newcommand{\psc}{\textsf{psc}}
\newcommand{\agg}{\textsf{agg}}
\newcommand{\prop}{\textsf{prop}}
\newcommand{\src}{\textsf{src}}
\newcommand{\sink}{\textsf{sink}}

\newcommand{\ie}{\textit{i.e.}}
\newcommand{\eg}{\textit{e.g.}}

\newcommand{\defpar}[1]{\smallskip\textbf{#1}}

\begin{document}
\title{Conquering the Extensional Scalability Problem for Value-Flow Analysis Frameworks}

\author{Qingkai Shi}
\orcid{1234-5678-9012}
\affiliation{%
  \institution{The Hong Kong University of Science and Technology}
  \city{Hong Kong} 
  \state{China}
}
\email{qshiaa@cse.ust.hk}

\author{Rongxin Wu}
\affiliation{%
  \institution{Xiamen University}
  \city{Xiamen} 
  \state{China}
}
\email{wurongxin@xmu.edu.cn}

\author{Gang Fan}
\affiliation{%
  \institution{The Hong Kong University of Science and Technology}
  \city{Hong Kong} 
  \state{China}
}
\email{gfan@cse.ust.hk}

\author{Charles Zhang}
\affiliation{%
  \institution{The Hong Kong University of Science and Technology}
  \city{Hong Kong} 
  \state{China}
}
\email{charlesz@cse.ust.hk}

\begin{abstract}

With an increasing number of value-flow properties to check,
existing static program analysis still tends to have scalability issues when high precision is required.
We observe that the key design flaw behind the scalability problem is that 
the core static analysis engine is oblivious of the mutual synergies among different properties being checked and,
thus, inevitably loses many optimization opportunities.
Our approach is inter-property-aware and
able to capture possible overlaps and inconsistencies among different properties.
Thus, before analyzing a program,
we can make optimization plans which decide how to reuse the specific analysis results of a property to speed up checking other properties.
Such a synergistic interaction among the properties significantly improves the analysis performance.
We have evaluated our approach
by checking twenty value-flow properties in standard benchmark programs and ten real-world software systems.
The results demonstrate that our approach is more than 8$\times$ faster than existing ones but consumes only 1/7 memory.
Such a substantial improvement in analysis efficiency is not achieved by sacrificing the effectiveness:
at the time of writing, \confirmedbugnum\ bugs found by our approach have been fixed by developers and four of them have been assigned CVE~IDs due to their security impact.

\end{abstract}

%
%



\maketitle

\section{Introduction}
\label{sec:intro}

Value flows~\cite{shi2018pinpoint,cherem2007practical,sui2016svf,livshits2003tracking}, 
which track how values are loaded and stored in the program, 
underpin the analysis for a broad category of software properties, 
such as memory safety (\eg, null dereference, double free, etc.), 
resource usage (\eg, memory leak, file usage, etc.), 
and security properties (\eg, the use of tainted data). 
In addition, there are a large and growing number of domain-specific value-flow properties. 
For instance, mobile software requires that the personal information cannot be passed to an untrusted code~\cite{arzt2014flowdroid}, 
and, in web applications, 
tainted database queries are not allowed to be executed~\cite{tripp2013andromeda}.
\fortify,\footnote{Fortify Static Analyzer: \url{https://www.microfocus.com/en-us/products/static-code-analysis-sast/}} a commercial static code analyzer, checks nearly ten thousand value-flow properties from hundreds of unique categories. 
Value flow problems exhibit a very high degree of versatility, 
which poses great challenges to the effectiveness of general-purpose program analysis tools. 

Faced with such a massive number of properties and the need of extension, 
existing approaches (\eg, \fortify, \clang\footnote{Clang Static Analyzer: \url{https://clang-analyzer.llvm.org/}} and \fbinfer\footnote{Infer Static Analyzer: \url{http://fbinfer.com/}}) 
provide a customizable framework together with a set of property interfaces that enable the quick customization for new properties. 
For instance, \clang\ uses a symbolic-execution engine such that, 
at every statement, 
it invokes the callback functions registered for the properties to check.
The callback functions 
are written by the framework users in order to
collect the symbolic-execution results, such as the symbolic memory and the path condition, 
so that we can judge the presence of any property violation at the statement. 
Despite the existence of many \clang-like frameworks,
when high precision like path-sensitivity is required, existing static analyzers still cannot scale well with respect to a large number of properties to check, which we refer to as the \textit{extensional scalability issue}. 
For example, our evaluation shows that
\clang\ cannot path-sensitively check twenty properties for many programs in ten hours. \pin~\cite{shi2018pinpoint} has already run out of 256GB memory for checking only eight properties.

We observe that, behind the extensional scalability issue,
the key design flaw in conventional extension mechanisms (like that in \clang)
is that the core static analysis engine is oblivious to the properties being checked.
Although the property obliviousness
gives the maximum flexibility and extensibility to the framework,
it also prevents
the core engine from utilizing the property-specific analysis results for optimization.
This scalability issue is slightly alleviated by a class of approaches that are property-aware and demand-driven~\cite{fan2019smoke,ball2002slam,le2008marple}. 
These techniques are scalable with respect to a small number of properties because the core engine can skip certain program statements by understanding what program states are relevant to the properties. 
However, in these approaches, the semantics of properties are also opaque to each other.
As a result, when the number of properties grows very large, the performance of the demand-driven approaches will quickly deteriorate, as in the case of \pin.
To the best of our knowledge,
the number of literature specifically addressing the extensional scalability issue is
very limited. Readers can refer to Section~\ref{sec:relatedwork} for a detailed discussion.

In this work, we advocate an inter-property-aware design to relax the property-property and the property-engine obliviousness so that 
the core static analysis engine can exploit the 
mutual synergies among different properties for optimization.
In our analysis,
such exploitation of mutual synergies are enabled 
by enforcing a simple value-flow-based property model,
which picks out source and sink values, respectively, as well as the predicate over these values for the satisfaction of the property. 
For instance, for a null deference property, our property model only requires the users of our framework to indicate where a null pointer may be created, where the null dereference may happen,
as well as a simple predicate that enforces the propagation of the null pointer.
Surprisingly, given a set of properties specified in our property model, 
our static analyzer can 
automatically understand the overlaps and inconsistencies of the properties to check.
Based on the understanding,
before analyzing a program, we can make dedicated analysis plans
so that, at runtime, the analyzer can 
transmit the analysis results on path-reachability and path-feasibility across different properties for optimization.
The optimization allows us to significantly reduce redundant graph traversals and unnecessary invocation of SMT solvers,
two critical performance bottlenecks of conventional approaches.
Section~\ref{sec:overview} provides examples to illustrate our approach.

We have implemented our approach, named \toolname, which is a new demand-driven and compositional static analyzer with the precision of path-sensitivity.
Like a conventional compositional analysis~\cite{xie2005scalable},
our implementation allows us to concurrently analyze functions that do not have calling relations.
In \toolname, we have included all C/C++ value-flow properties that \clang\ checks by default.
In the evaluation, we compare \toolname\ with three state-of-the-art bug-finding tools, \pin, \clang, and \fbinfer, using a standard benchmark and ten popular industrial-sized software systems.
The experimental results demonstrate that \toolname\ is more than 8$\times$ faster than \pin\ but consumes only 1/7 memory.
It is as efficient as \clang\ and \fbinfer\ in terms of both time and memory cost but is much more precise.
Such promising scalability of \toolname\ is not achieved by sacrificing the capability of bug finding.
In our experiments, although the benchmark software systems have been checked by numerous free and commercial
tools,
\toolname\ is still able to detect many previously-unknown bugs, in which \confirmedbugnum\ have been fixed by the developers and four have been assigned CVE~IDs due to their security impact.

In summary, our main contributions are listed as following:
\begin{itemize}
	\item An inter-property-aware design for checking value-flow properties, which mitigates the extensional scalability issue.
	
	\item A series of cross-property optimization rules that can be made use of for general value-flow analysis frameworks.
	
	\item A detailed implementation and a systematic evaluation that demonstrates our high scalability, precision, and recall.
\end{itemize}

\section{Overview}\label{sec:overview}

The key factor that allows us to conquer the extensional scalability problem
is the exploitation of the mutual synergies among different properties.
In this section,
we first use two simple examples to illustrate the mutual synergies
and then provide a running example
used in the whole paper.

\subsection{Mutual Synergies}

We observe that the mutual synergies among different properties are 
originated from their overlaps and inconsistencies.

\begin{figure}[t]
	\centering
	\includegraphics[width=\columnwidth]{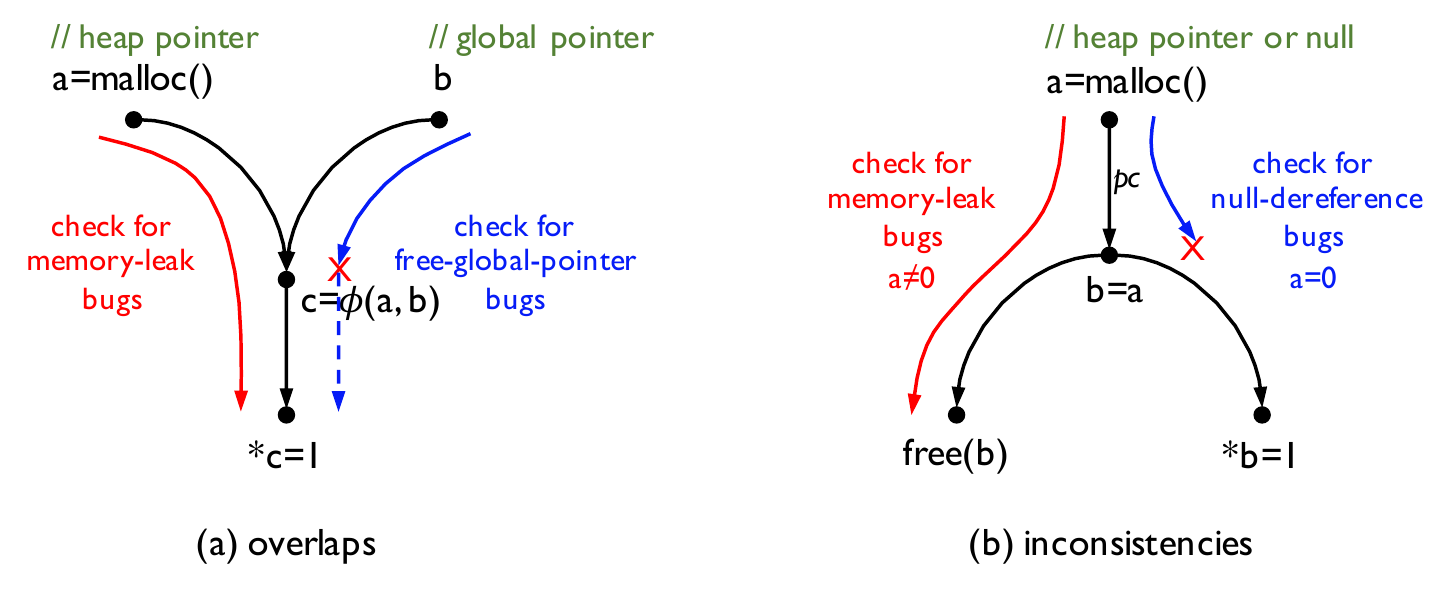}
	\caption{Possible overlaps and inconsistencies among properties. Each edge represents a value flow.}
	\label{fig:intro_insight}
\end{figure}

In Figure~\ref{fig:intro_insight}a,
to check memory-leak bugs,
we need to track value flows from the newly-created heap pointer $a$ to check if the pointer will be freed.\footnote{In the paper, a pointer $p$ is ``freed'' means it is used in the function call \textit{free(p)}. We will detail how to use the value-flow information to check bugs later.}
To check free-global-pointer bugs,
we track value flows from the global variable $b$ to check if it will be freed.\footnote{Freeing a pointer pointing to memory not on the heap (e.g., memory allocated by global variables) is buggy. See details in \url{https://cwe.mitre.org/data/definitions/590.html}}
As illustrated, the value-flow paths to search overlap from \textit{c=$\phi$(a,b)} to \textit{*c=}1.
Being aware of such overlaps,
when traversing the graph from \textit{a=malloc()} for memory-leak bugs,
we can record that \textit{c=$\phi$(a,b)} cannot reach any free operation.
Then, when checking free-global-pointer bugs,
we can use this recorded information to immediately stop the graph traversal at the vertex \textit{c=$\phi$(a,b)}, thereby saving computation resources.

In Figure~\ref{fig:intro_insight}b,
to check memory-leak bugs, we track value flows from the newly-created pointer $a$ to where it is freed.
To check null-dereference bugs, 
considering that \textit{malloc()} may return a null pointer when memory allocation fails,
we track value flows from the same pointer $a$ to where it is dereferenced.
The two properties have an inconsistent constraint:
the former requires \textit{a}$\ne$0 so that $a$ is a valid heap pointer
while the latter requires \textit{a=}0 so that $a$ is a null pointer.
Being aware of the inconsistency,
when traversing the graph for checking null dereferences,
we can record whether \textit{pc} (the path condition from \textit{a=malloc()} to \textit{b=a}) and \textit{pc}$\land$\textit{a}=0 can be satisfied.
If \textit{pc} can be satisfied but \textit{pc}$\land$\textit{a}=0 not,
we can confirm that \textit{pc}$\land$\textit{a}$\ne$0 must be satisfiable without an expensive constraint-solving procedure,
thus speeding up the process of checking memory leaks.

\subsection{A Running Example}
\label{subsec:running}

\begin{figure}[t]
	\centering
	\includegraphics[width=\columnwidth]{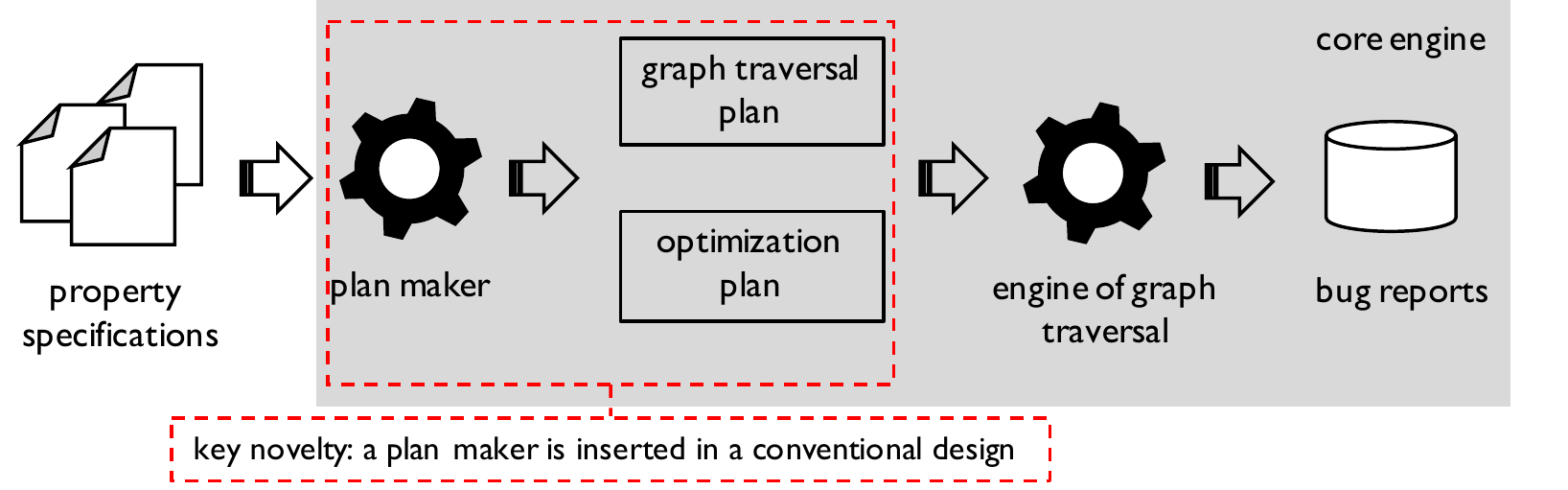}
	\caption{The workflow of our approach.}
	\label{fig:sec2_arch}
\end{figure}

\begin{figure}[t]
	\centering
	\includegraphics[width=\columnwidth]{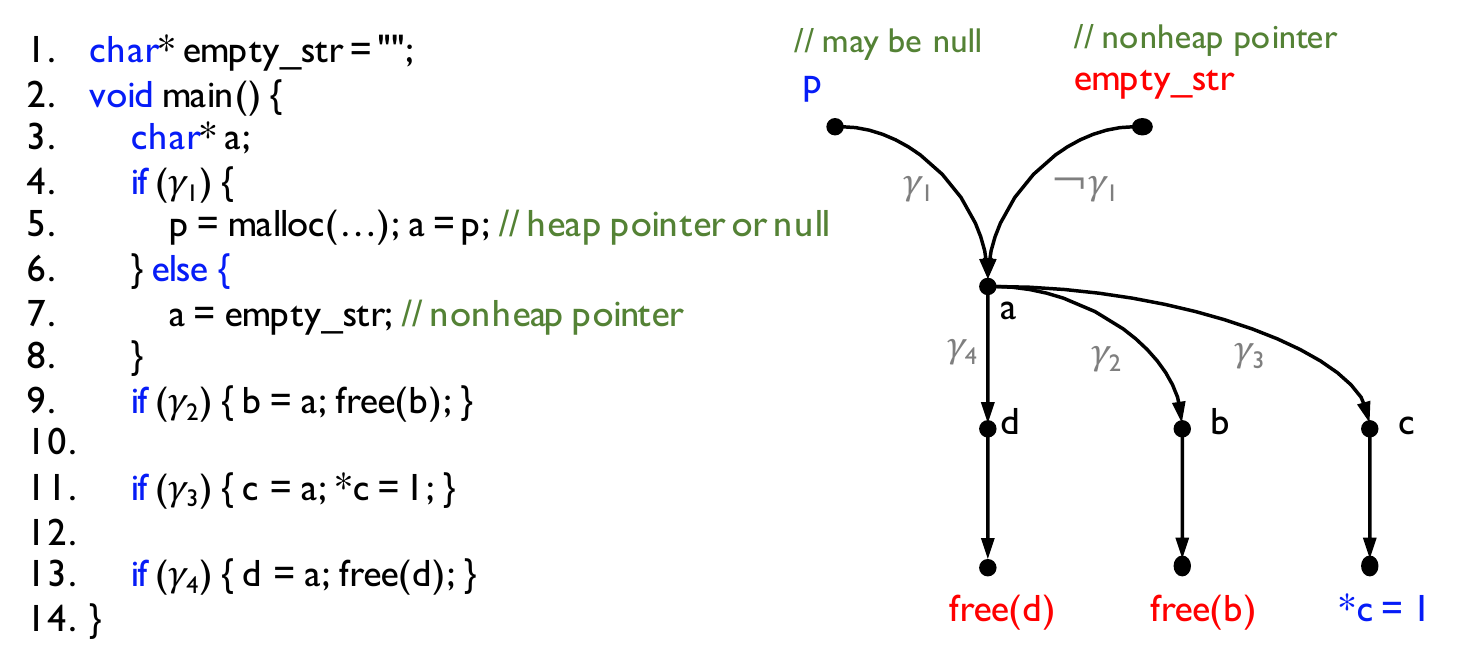}
	\caption{An example to illustrate our method.}
	\label{fig:sec2_example}
\end{figure}

Let us describe a running example using the value-flow graph in Figure~\ref{fig:sec2_example},
where we check null-deference and free-global-pointer bugs following the workflow in Figure~\ref{fig:sec2_arch}.
Given the program,
we firstly follow previous works to build the value-flow graph~\cite{sui2014detecting,cherem2007practical,shi2018pinpoint}.
With the graph in hand,
we check the two properties with the precision of path-sensitivity.
Here, path-sensitivity means that when searching paths on the value-flow graph,
we will invoke an SMT solver to solve path conditions and other property-specific constraints,
so that infeasible paths are pruned.

\defpar{The Property Specifications.}
As is common practice,
users of our framework need to provide the property specifications.
The users are responsible for the correctness of the specifications.

In this paper, we focus on value-flow properties,
which are checked by examining a series of value-flow paths from certain source values to some sink values.
As an overview,
the specifications of the two properties are described as two quadruples:
$$
\prop\ \textit{null-deref} := (v = \textit{malloc}(\_); \_ = *v, *v = \_; v = 0; \textsf{never})
$$
$$
\prop\ \textit{free-glob-ptr} := (\textit{glob}; \textit{free}(v); \textit{true}; \textsf{never})
$$
As illustrated above,
the specification of a value-flow property consists of four parts which are separated by the semicolons.
The first and second parts are the source and sink values.
The values are specified by pattern expressions, which represent the values at certain statements.
The uninterested values are written as ``\_''.
In the example,
the source values of \textit{null-deref} and \textit{free-glob-ptr} are the return pointer of \textit{malloc()} and the global pointer \textit{empty\_str}, respectively.
The sink value of \textit{null-deref} is the dereferenced value $c$ at the statement \textit{*c}=1
The sink values of \textit{free-glob-ptr} are the freed values at \textit{free(b)} and \textit{free(d)}.

The third part is a property-specific constraint, which is the precondition on which the bug can happen.
The constraint of \textit{null-deref} is to require the value on a value-flow path to be a null pointer, \ie, $v=0$.
The property \textit{free-glob-ptr} does not have any specific constraint and, thus, puts \textit{true} in the quadruple.

The predicate ``\textsf{never}'' means that value-flow paths between the specified sources and sinks should never be feasible. Otherwise, a bug exists.

\defpar{The Core Static Analysis Engine.}
Before the analysis,
our core engine automatically makes analysis plans based on the specifications.
The analysis plans include the graph traversal plan and the optimization plan.
In the example,
we make the following optimization plans:
(1) checking \textit{free-glob-ptr} before \textit{null-deref};
(2) when traversing the graph for checking \textit{free-glob-ptr},
we record the vertices that cannot reach any sink vertex of \textit{null-deref}.
The graph traversal plan in the example is trivial,
which is to traverse the graph from each source vertex of each property.

In Figure~\ref{fig:sec2_example}, when traversing the graph from \textit{empty\_str} to check \textit{free-glob-ptr},
the core engine will visit all vertices except $p$ to look for free operations.
According to the optimization plan,
during the graph traversal,
the core engine records that $b$ and $d$ cannot reach any dereference operation.

For \textit{null-deref},
we traverse the graph from \textit{p}.
When visiting $b$ and $d$, 
since the previously-recorded information tells that they cannot reach any sink vertex,
we prune the subsequent paths from $b$ and $d$ and
only need to continue the graph traversal from $c$.

It is noteworthy that if we check \textit{null-deref} before \textit{free-glob-ptr},
we only can prune one path from $c$ for \textit{free-glob-ptr} based on the results of \textit{null-deref} (see Section~\ref{subsubsec:opt}).
We will further explain the rationale of our analysis plans in the following sections.

\section{Value-Flow Properties}
\label{sec:preliminaries}


This section provides a specification model for value-flow properties with the following two motivations.
On the one hand, 
we observe that many property-specific constraints play a significant role in performance optimization.
The specific constraints of a property not only can be used to optimize the property itself,
but also can benefit other properties being checked together.
Existing value-flow analyses either ignore or do not well utilize property-specific constraints,
which exacerbates the extensional scalability issue.

On the other hand, despite many studies on value-flow analysis~\cite{livshits2003tracking,shi2018pinpoint,sui2014detecting,sui2016svf,cherem2007practical}, 
we are still lack of a general and extensible specification model that can widen the opportunities of sharing analysis results across different properties.
Some of the existing studies only focus on checking a specific property (\eg, memory leak~\cite{sui2014detecting}).
Some adopt different specifications to check the same value-flow property (\eg, double free~\cite{shi2018pinpoint, cherem2007practical}).

\defpar{Preliminaries.}
As existing works~\cite{li2011boosting,sui2014detecting,shi2018pinpoint},
we assume that the code in a program is in static single assignment (SSA) form,
where every variable has only one definition~\cite{cytron1991efficiently}.
Also,
we say the value of a variable $a$ flows to a variable $b$ (or $b$ is data-dependent on $a$) if 
$a$ is assigned to $b$ directly (via assignments, such as \textit{b=a}) or indirectly (via pointer dereferences, such as \textit{*p=a; q=p; b=*q}).
Thus, a value-flow graph can be defined as a directed graph
where the vertices are values in the program
and the edges represent the value-flow relations.
A path is called value-flow path if it is a path on the value-flow graph.

\defpar{Property Specification.} As defined below, we model a value-flow property as an aggregation of value-flow paths.

\begin{definition}[Value-Flow Property]
	A value-flow property, $x$, is a quadruple: $\prop\ x := (\src; \sink; \psc; \agg)$, where
	\begin{itemize}
		\item 
		$\src$ and $\sink$ are two pattern expressions (Table~\ref{tab:pattern}) that specify the source and sink values of the value-flow paths to track.
		
		\item
		$\textsf{psc}$ is
		the property-specific constraint that every value on the value-flow path needs to satisfy.
		
		\item 
		$\textsf{agg} \in \{\textsf{never}, \textsf{must}, \textsf{never-sim}, \cdots \}$ is an extensible predicate 
		that determines how to aggregate value-flow paths to check the specified property.
	\end{itemize}
	\label{def:vuln}
\end{definition}

\begin{table}
	\small
	\renewcommand\arraystretch{0.8}
	\centering
	\caption{Pattern expressions used in the specification}
	\label{tab:pattern}
	\begin{tabular}{|p{2mm}lp{33mm}lp{2mm}|}
		\hline
		&&&&\\
	&	$p$ & $::=$                             & ::~\textbf{patterns}     &\\
	&	    & | $p_1, p_2, \cdots$              & ::~\textbf{pattern list} &\\
	&	    & | $v_0 = \textit{sig}(v_1, v_2, \cdots)$   & ::~\textbf{call}         &\\
	&	    & | $v_0 = *v_1$                    & ::~\textbf{load}         &\\
	&	    & | $*v_0 = v_1$                    & ::~\textbf{store}        &\\
	&	    & | $v_0 = v_1$                     & ::~\textbf{assign}       &\\
	&	    & | \textit{glob}                   & ::~\textbf{globals}       &\\
	&	$v$ & $::=$                             & ::~\textbf{symbol}      &\\
	&	    & | \textit{sig}                    & ::~\textbf{character string}      &\\
	&	    & | \_                              & ::~\textbf{uninterested value} &\\
	&&&&\\\hline
	&&&&\\
	Examples: &&&&\\
	&&$v = \textit{malloc}(\_)$ & ret values of any state-         &\\
	&&&ment calling \textit{malloc};& \\
	&&$\_ = \textit{send}(\_,v,\_,\_)$ & the 2nd arg of any sta-          &\\
	&&&tement calling \textit{send};& \\
	&&{$\_ = *v$}                 & dereferenced values at &\\
	&&&every load statement;& \\
	&&&&\\\hline
	\end{tabular}
\end{table}

In practice, we can use the quadruple to specify a wide range of value-flow properties.
As discussed below, we put the properties into three categories,
which are checked by aggregating a single, two, or more value-flow paths, respectively.

\defpar{Null-Dereference-Like Bugs.}
Many program properties can be checked using a single value-flow path,
such as \textit{null-deref} and \textit{free-glob-ptr} defined in Section~\ref{subsec:running}, as well as
a broad range of taint issues that 
propagate a tainted object to a program point consuming the object~\cite{denning1976lattice}.

\defpar{Double-Free-Like Bugs.}
A wide range of bugs happen in a program execution because two program statements (\eg, two statements calling function \textit{free}) consecutively operate on the same value (\eg, a heap pointer).
Typical examples include use-after-free which is a general form of double free, as well as bugs that operate on expired resources such as using a closed file descriptor.
As an example, 
the specification of double-free can be specified as
$$
\prop\ \textit{double-free} := (v = \textit{malloc}(\_); \textit{free}(v); v \ne 0; \textsf{never-sim})
$$

In the specification, the property-specific constraint $v\ne0$ requires the initial value (or equivalently, all values) on the value-flow path is a valid heap pointer.
This is because $v=0$ means \textit{malloc()} fails to allocate memory but returns a null pointer. In this case, the free operation is harmless.
The aggregate predicate ``\textsf{never-sim}'' means that the value-flow paths from the same pointer should never occur simultaneously. 
In other words, there is no control-flow path that goes through two different free operations on the same heap pointer.
Otherwise, a double-free bug exists.

In Figure \ref{fig:sec2_example}, 
for the two value-flow paths from $p$ to the two \textit{free} operations,
we can check $(\gamma_1\land \gamma_2) \land (\gamma_1\land \gamma_4) \land (p\ne0)$ to check double-free bugs.
Here,
$(\gamma_1\land \gamma_2)$ and $(\gamma_1\land \gamma_4)$ are the path conditions of the two paths, respectively.

\defpar{Memory-Leak-Like Bugs.}
Many bugs happen because a value (\eg, a heap pointer) must be properly handled (\eg, freed by calling function \textit{free}) in any program execution but, unfortunately, not.
Typical examples include all kinds of resource leaks such as file descriptor leak, internet socket leak, etc.
As an example,
we write the following specification for checking memory leaks:
$$
\prop\ \textit{mem-leak} := (v = \textit{malloc}(\_); \textit{free}(v); v \ne 0; \textsf{must})
$$

Compared to \textit{double-free},
the only difference is the aggregate predicate.
The aggregate predicate ``\textsf{must}'' means that the value-flow path from a heap pointer 
must be able to reach a free operation. Otherwise, a memory leak exists in the program.

In Figure~\ref{fig:sec2_example}, for the two value-flow paths from $p$ to the two \textit{free} operations,
we can check the disjunction of their path conditions, \ie,
$\lnot ((\gamma_1\land \gamma_2)\lor (\gamma_1\land \gamma_4)) \land \gamma_1 \land (p\ne0)$ to determine if a memory leak exists.
Here,
$(\gamma_1\land \gamma_2)$ and $(\gamma_1\land \gamma_4)$ are the path conditions of the two paths, respectively.
The additional $\gamma_1$ is the condition on which the heap pointer is created.
\section{Inter-property-aware analysis}
\label{sec:core}

Given multiple value-flow properties specified as the quadruple $(\src; \sink; \textsf{psc}; \textsf{agg})$,
our inter-property-aware static analyzer then starts to check them by searching value-flow paths and finally checking bugs based on the \textsf{agg} predicate.
Since the path aggregate step is easy to run in parallel by independently checking all possible path groups, it is not the performance bottleneck.
In this paper,
we concentrate on how to exploit mutual synergies among different properties to improve the efficiency of searching value-flow paths.


\subsection{A Na\"ive Static Analyzer}
\label{subsec:naive}

\begin{algorithm}[t]
	\KwIn{the value-flow graph of a progam to check}
	\KwIn{a set of value-flow properties to check}
	\KwOut{paths between sources and sinks for each property}
	
	\ForEach{property in the input property set}{
		\ForEach{source $v$ in its source set}{
			\While{visit $v'$ in the depth-first search from $v$}{
				\colorbox[gray]{0.95}{
					\begin{minipage}{6cm}
						\If{$\psc$ cannot be satisfied}{
							stop the search from $v'$\;
						}
					\end{minipage}
				}
			}
		}
	}
	\caption{The na\"ive static analyzer.}
	\label{alg:naive}
\end{algorithm}

For multiple value-flow properties,
a na\"ive static analyzer checks them independently in a demand-driven manner.
As illustrated in Algorithm~\ref{alg:naive},
for each value-flow property,
the static analyzer traverses the value-flow graph from each of the source vertices.
At each step of the graph traversal,
we check if $\textsf{psc}$ can be satisfied with regard to the current path condition.
If not,
we can stop the graph traversal along the current path to save computing resources.
This path-pruning process is illustrated in the shaded part of Algorithm~\ref{alg:naive},
which is a critical factor to improve the analysis performance.

We observe that the properties to check usually have overlaps and inconsistencies
and, thus, are not necessary to be checked independently as the na\"ive approach.
Instead,
we can exploit 
the overlaps and inconsistencies to facilitate the path-pruning process in Algorithm~\ref{alg:naive},
thus improving the analysis efficiency.
In what follows,
we detail how the mutual synergies are utilized.

\begin{table*}[t]
	\centering
	\caption{Rules of Making Analysis Plans for a Pair of Properties}
	\label{tab:plans}
	\begin{tabular}{c|l|l|l|l}
		\toprule
		\multicolumn{5}{c}{\textbf{Optimization Plans}} \\
		\multicolumn{5}{c}{$\prop\ x :=(\src_1;\sink_1;\psc_1;\agg_1)$ and $\prop\ y :=(\src_2; \sink_2;\psc_2;\agg_2)$, $\src_1\ne\src_2$} \\
		\midrule
		
		\textbf{ID} & \textbf{Rule Name}  & \textbf{Precondition}  & \textbf{Plan}   & \textbf{Benefit}     \\
		\midrule
		
		1           & property ordering        & $\#\sink_1 > \#\sink_2$    & check $x$ before $y$     & more chances to prune paths \\\hline
		
		2           & \multirow{3}{*}{result recording}                    & check $x$ before $y$                      & record vertices that cannot reach $\sink_2$            & prune paths at a vertex                                         \\\cline{1-1}\cline{3-5}
		3           &                                                      & check $x$ before $y$, $\psc_1 = \psc_2$   & record unsat cores that conflict with $\psc_2$         & prune paths if going through \\\cline{1-1}\cline{3-4}
		4           &                                                      & check $x$ before $y$, $\psc_1 \ne \psc_2$ & record interpolants that conflict with $\psc_2$        &      a set of edges        \\
		
		\bottomrule
\multicolumn{5}{c}{ }\\
		\toprule
		\multicolumn{5}{c}{\textbf{Graph Traversal Plans}}   \\
		\multicolumn{5}{c}{$\prop\ x :=(\src_1;\sink_1;\psc_1;\agg_1)$ and $\prop\ y :=(\src_2; \sink_2;\psc_2;\agg_2)$, $\src_1=\src_2$} \\
		\midrule
		
		\textbf{ID} & \textbf{Rule Name}  & \textbf{Precondition}  & \textbf{Plan}   & \textbf{Benefit}     \\
		\midrule
		
		5           & traversal merging                                    & -                                         & search from $\src_1$ for both properties & sharing path conditions                                         \\\hline

		6           & \multirow{5}{*}{\psc -check ordering} & $\psc_1 \land \psc_2 = \psc_1$            & check $\psc_1$ first                                   & if satisfiable, so is $\psc_2$                                  \\\cline{1-1}\cline{3-5}
		
		\multirow{2}{*}{7}  &   & \multirow{2}{*}{$\psc_1 \land \psc_2 \ne \textit{false}$}  & \multirow{2}{*}{check $\psc_1 \land \psc_2$}                            & if satisfiable, both $\psc_1$ and    \\
		&&&&  $\psc_2$ can be satisfied     \\\cline{1-1}\cline{3-5}
		
		\multirow{2}{*}{8}  &   & \multirow{2}{*}{$\psc_1 \land \psc_2 = \textit{false}$}  & \multirow{2}{*}{check any, \eg, $\psc_1$, first}                            & if unsatisfiable, $\psc_2$ can be \\
		&&&&   satisfied     \\
		
		\bottomrule
	\end{tabular}
\end{table*}

\subsection{Optimized Intra-procedural Analysis}
\label{subsec:local}

Based on the input property specifications,
the core static analysis engine makes two plans for traversing the value-flow graph.
The first is the optimization plan, which aims to prune more paths than the na\"ive  approach.
The second is the graph traversal plan, which concerns how to share paths among properties rather than prune paths.
As a whole, all the plans are summarized in Table~\ref{tab:plans}.
Each row of the table 
is a rule describing what plan we can make on certain preconditions and what benefits we can obtain from the plan.
To be clear, in this section, we detail the plans in the context of scanning a single-procedure program.
In the next subsection, we introduce the inter-procedural analysis.

\subsubsection{Optimization Plan}
\label{subsubsec:opt}

Based on the property specifications,
we adopt several strategies to facilitate the path pruning (Rules 1 -- 4 in Table~\ref{tab:plans}).

\defpar{Ordering the Properties (Rule 1).}
Given a set of properties with different source values,
we need to determine the order in which they are checked.
Generally,
there is no perfect order that can guarantee the best optimization results.
However,
we observe that a random order could significantly affect how many paths we can prune in practice.

Let us consider the example in Figure~\ref{fig:sec2_example} again.
In Section~\ref{subsec:running},
we have explained that 
if \textit{free-glob-ptr} is checked before \textit{null-deref},
we can prune the two paths from $b$ and $d$ when checking \textit{null-deref}.
However, if we change the checking order, \ie, check \textit{null-deref} before \textit{free-glob-ptr},
we can only prune one path from $c$.
In detail,
when checking \textit{null-deref},
the core engine records that $c$ cannot reach any sinks of \textit{free-glob-ptr}.
In this case, we can prune the path from $c$ when checking \textit{free-glob-ptr}.

Intuitively, what makes the number of pruned paths different 
is that the number of free operations is more than dereference operations in the value-flow graph.
That is, the more sink vertices in the value-flow graph, the fewer paths we can prune for the property.
Inspired by this intuition and the example,
the order of property checking is arranged according to the number of sink vertices.
That is, the more sink vertices in the value-flow graph, the earlier we check this property.

\defpar{Recording Sink-Reachability (Rule 2).}
Given a set of properties $\{ \prop_1$, $\prop_2, \cdots \}$,
the basic idea is that,
when checking $\prop_i$ by traversing the value-flow graph, 
the core engine needs to record whether each visited vertex may reach a sink vertex of $\prop_j (j\ne i)$.
With the recorded information,
when checking $\prop_j (j\ne i)$ and visiting a vertex that cannot reach any of its sinks,
the path from the vertex can be pruned.
Section~\ref{subsec:running} illustrates the method.

\defpar{Recording the \psc-Check Results (Rules 3 \& 4).}
Given a set of properties $\{ \prop_1$, $\prop_2, \cdots \}$,
the basic idea is that,
when checking $\prop_i$ by traversing the value-flow graph, 
the core engine needs to record whether some path segments (or a set of edges) conflict with the property-specific constraint $\psc_j$ of $\prop_j (j\ne i)$.
With the recorded information,
when checking $\prop_j (j\ne i)$ and visiting the path segments that do not satisfy its specific constraint,
the path with this segment can be pruned.

Let us consider the running example in Figure~\ref{fig:sec2_example} again
and assume that $\gamma_3$ is $a\ne0$.
When traversing the graph from \textit{empty\_str} to check \textit{free-glob-ptr},
the core engine needs to record that 
the edge from $a$ to $c$ (whose condition is $\gamma_3$, \ie, $a\ne0$) conflicts with the property-specific constraint of \textit{null-deref} (\ie, $a=0$).
With this information,
when checking \textit{null-deref} by traversing the graph from $p$,
we can also prune the path from the edge.

In practice,
although the property-specific constraints are usually simple,
the path constraints, \eg, $\gamma_3$ in the above example, are usually very sophisticated.
Fortunately,
thanks to the advances in the area of clause learning~\cite{beanie2003understanding},
we are able to efficiently compute some reusable facts when using SMT solvers to check path conditions and property-specific constraints.
Specifically,
we compute two reusable facts when a property-specific constraint $\psc_i$ conflicts with the current path condition $\pc$.

When $\pc\land \psc_i$ is unsatisfiable,
we can record the unsatisfiable core~\cite{dershowitz2006scalable},
which is a set of Boolean predicates from $\pc$, \eg, $\{ \gamma_1, \gamma_2, \cdots\}$, 
such that $\gamma_1 \land \gamma_2 \land \cdots \land \psc_i = \textit{false}$.
Since $\pc$ is the conjunction of the edge constraints on the value-flow path,
each $\gamma_i$ corresponds to the condition of an edge $\epsilon_i$ on the value-flow graph.
Thus, 
we can record an edge set $\{ \epsilon_1, \epsilon_2, \cdots\}$,
which conflicts with $\psc_i$.
When checking the other property with the same property-specific constraint,
if a value-flow path goes through these recorded edges, we can prune the remaining paths.

In addition to the unsatisfiable cores,
we also can record the interpolation constraints~\cite{cimatti2010efficient}, 
which are even reusable for properties with a different property-specific constraint.
In the above example,
assume that $\psc_i$ is $a=0$ and $\{ \gamma_1, \gamma_2, \cdots\}$ is $\{ a+b > 3, b<0 \}$.
During the constraint solving, an SMT solver can refute the satisfiability of $(a+b > 3)\land (b<0) \land (a=0)$ by finding an interpolant $\gamma'$ such that $(a+b > 3)\land (b<0)\Rightarrow \gamma'$ but $\gamma'\not\Rightarrow (a=0)$.
In the example, the interpolant $\gamma'$ is $a>3$, which
provides a detailed explanation why the $\gamma$ set conflicts with $a=0$.
In addition,
the interpolant also indicates that the $\gamma$ set conflicts with many other constraints like $a < 0$, $a < 3$, etc.
Thus, given a property whose specific constraint conflicts with the interpolation constraint, 
it is sufficient to conclude that any value-flow path passing through the edge set can be pruned.

\subsubsection{Graph Traversal Plan}

Different from the optimization plan that aims to prune paths,
the graph traversal plan is to provide strategies to share paths among different properties.

\defpar{Merging the Graph Traversal (Rule 5).}
We observe that many properties actually share the same or a part of source vertices and even the same sink vertices.
If the core engine checks each property one by one,
it will inevitably repeat traversing the graph from a source vertex for different properties.
To avoid such repetitive graph traversal from the same source,
we propose the graph traversal plan to merge the path searching processes for different properties.

\begin{figure}[t]
	\centering
	\includegraphics[width=\columnwidth]{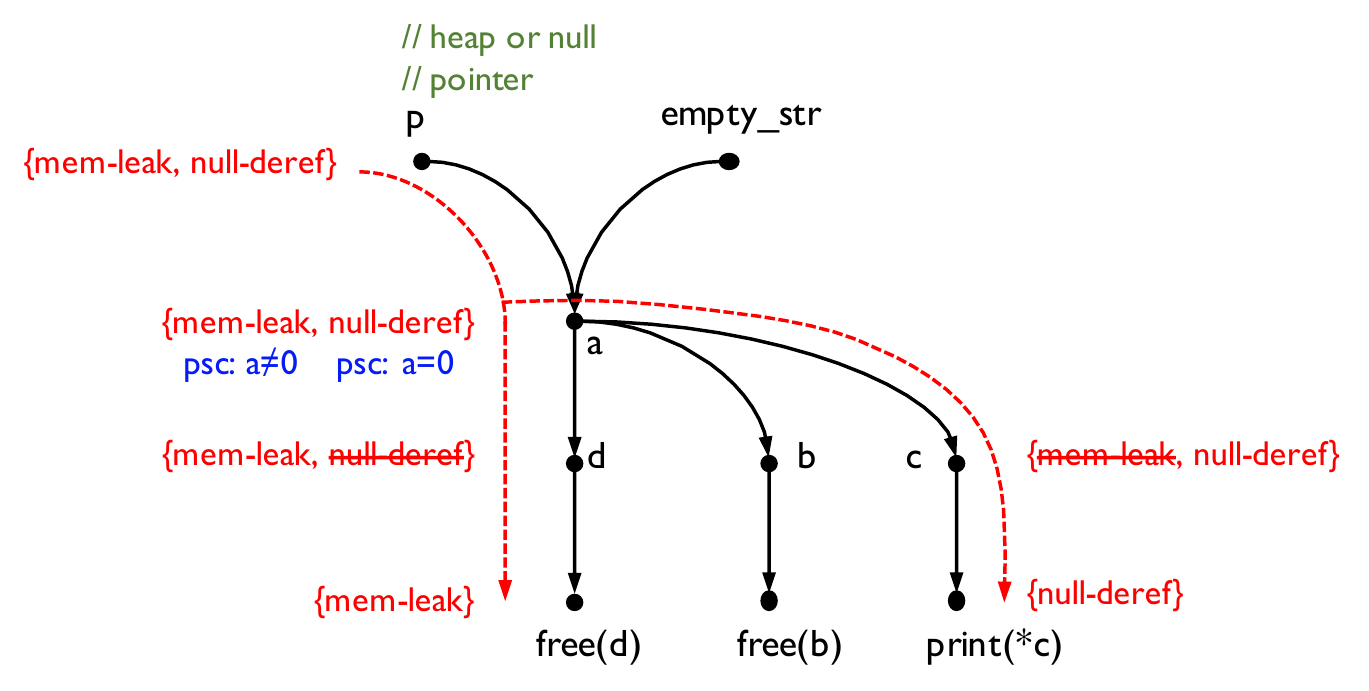}
	\caption{Merging the graph traversal.}
	\label{fig:core_merge}
\end{figure}

As an example, in Figure~\ref{fig:sec2_example},
since $p$ may be a heap pointer or null,
checking both \textit{null-deref} and \textit{mem-leak} needs to traverse the graph from $p$.
Figure~\ref{fig:core_merge} illustrates how the merged traversal is performed.
That is, we maintain a property set during the graph traversal to record what properties the current path contributes to.
Whenever visiting a vertex,
we check if a property needs to be removed from the property set.
For instance, at the vertex $d$, we may be able to remove \textit{null-deref} from the property set if we can determine $d$ cannot reach any dereference operation.
When the property set becomes empty at a vertex, the graph traversal stops immediately.

\defpar{Ordering the \psc-Checks (Rules 6 -- 8).}
Since the graph traversals are merged for different properties,
at a vertex, \eg, $a$ in Figure~\ref{fig:core_merge},
we have to check multiple property-specific constraints,
\eg, $a\ne0$ for \textit{mem-leak} and $a=0$ for \textit{null-deref}.
Thus,
a problem we need to address is to determine the order in which the property-specific constraints are checked.
Since checking such constraints often needs expensive SMT solving procedures,
the order of such constraint solving affects the analysis performance.

Given two property-specific constraints $\psc_1$ and $\psc_2$ as well as the current path condition $\pc$,
we consider three cases, \ie, $\psc_1 \land \psc_2 = \psc_1$, $\psc_1 \land \psc_2 \ne \textit{false}$,
and $\psc_1 \land \psc_2 = \textit{false}$, as listed in Table~\ref{tab:plans}.
Since property-specific constraints are usually simple,
the above relations between $\psc_1$ and $\psc_2$ are easy to compute. 

First, if $\psc_1 \land \psc_2 = \psc_1$, 
it means that the solution of $\psc_1$ also satisfies $\psc_2$.
Thus, we check $\pc \land \psc_1$ first. If it is satisfiable, 
we can confirm that $\pc \land \psc_2$ must be satisfiable without an expensive SMT solving procedure.

Second, if $\psc_1 \land \psc_2 \ne \textit{false}$, 
it means that there exists a solution that satisfying both $\psc_1$ and $\psc_2$.
In this case,
we check $\pc \land \psc_1 \land \psc_2$ first,
if it is satisfiable,
we can confirm both $\pc \land \psc_1$ and $\pc \land \psc_2$ can be satisfied
without additional SMT solving procedures.
In our experience, this strategy saves a lot of resources.

Third, if $\psc_1 \land \psc_2 = \textit{false}$, 
it means that there does not exist any solution that satisfies both $\psc_1$ and $\psc_2$.
In this case,
we check any, \eg, $\pc \land \psc_1$, first.
If the current path is feasible but $\pc \land \psc_1$ is not satisfiable,
we can confirm that $\pc \land \psc_2$ can be satisfied without invoking SMT solvers.
This case was illustrated in Figure~\ref{fig:intro_insight}b.

\subsection{Modular Inter-procedural Analysis}
\label{subsec:modular}

Scalable program analyses work by exploiting the modular structure of programs.
Almost every inter-procedural analysis builds
summaries for functions and reuses the function summary at its calling contexts, in order to scale to large programs~\cite{cousot2002modular,xie2005scalable}.
In \toolname,
we can seamlessly extend our optimized intra-procedural analysis to modular inter-procedural analysis
by exploring the local value-flow graph of each function and then stitching the local paths together to generate complete value-flow paths.
In the following, we explain our design of the function summaries.

\begin{figure}[t]
	\centering
	\includegraphics[width=\columnwidth]{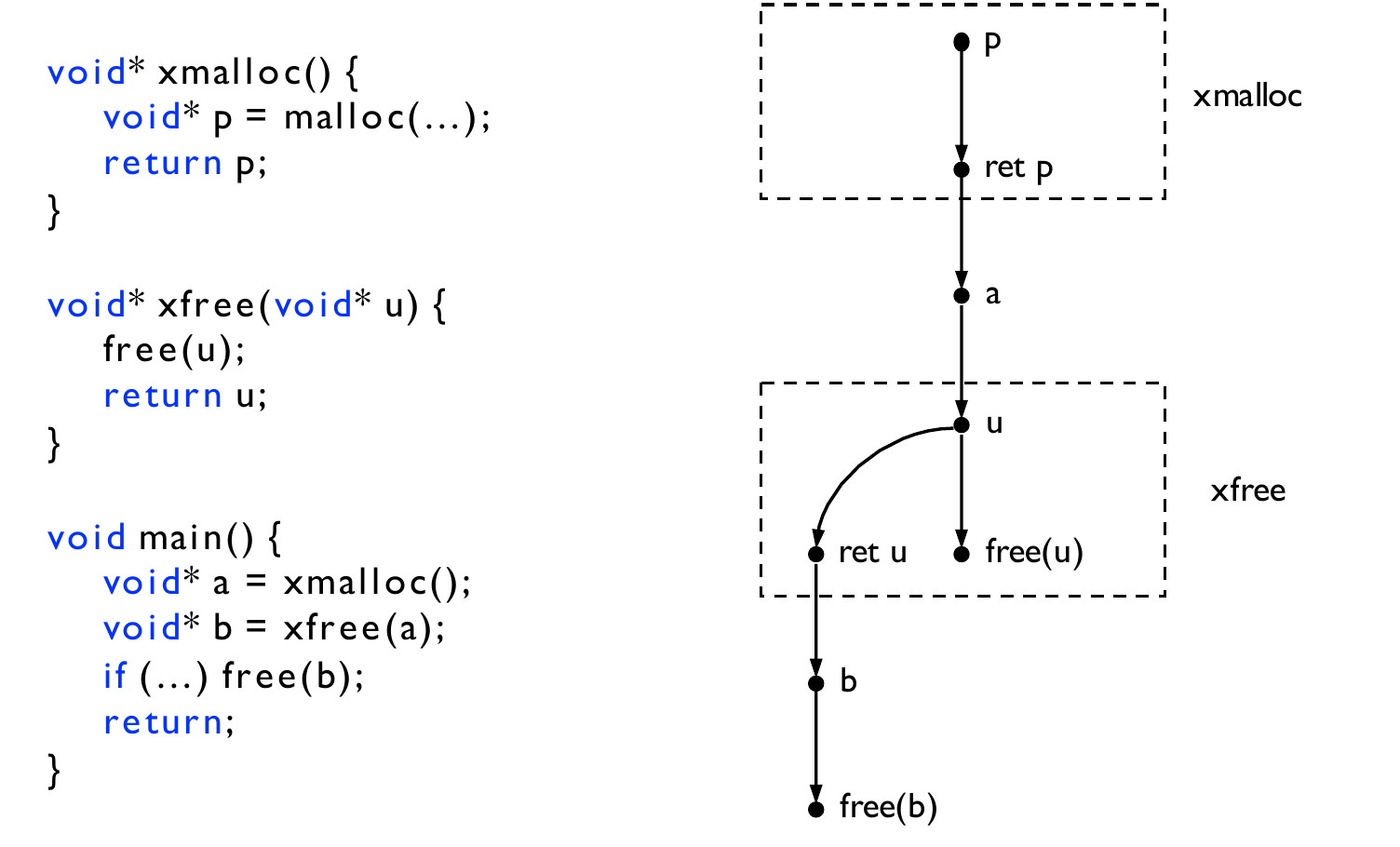}
	\caption{An example to show the inter-procedural analysis.}
	\label{fig:core_example}
\end{figure}

In our analysis, for each function, we build three kinds of value-flow paths as the function summaries.
They are defined as below and, 
in Appendices A and B,
we formally prove the sufficiency to generate these function summaries.
Intuitively, these summaries describe how function boundaries (i.e., formal parameters and return values)
partition a complete value-flow path.
Using the property \textit{double-free} as an example, 
a complete value-flow path from $p$ to \textit{free(b)} in Figure~\ref{fig:core_example}
is partitioned to a sub-path from $p$ to \textit{ret p} by the boundary of \textit{xmalloc()}.
This sub-path is an output summary of \textit{xmalloc()} as defined below.

\begin{definition}[Transfer Summary]
	\label{xdef:sl}
	Given a function \textit{f},
	a transfer summary of \textit{f}
	is a value-flow path from one of its formal parameters to one of its return values.
\end{definition}

\begin{definition}[Input Summary]
	\label{xdef:in}
	Given a function \textit{f},
	an input summary of \textit{f}
	is a value-flow path from one of its formal parameters to a sink value in \textit{f} or in the callees of \textit{f}.
\end{definition}

\begin{definition}[Output Summary]
	\label{xdef:out}
	Given a function \textit{f},
	an output summary of \textit{f}
	is a value-flow path from a source value to one of \textit{f}'s return values.
	The source value is in \textit{f} or in the callees of \textit{f}.
\end{definition}

After generating the function summaries, to avoid separately storing them for different properties,
each function summary is labeled with a bit vector to record what properties it is built for.
Assume that we need to check \textit{null-deref}, \textit{double-free}, and \textit{mem-leak} in Figure \ref{fig:core_example}.
The three properties are assigned with three bit vectors $0b001$, $0b010$, and $0b100$ as their identities, respectively.
As explained before, 
all three properties regard $p$ as the source vertex.
The sink vertices for checking \textit{double-free} and \textit{mem-leak} are \textit{free(b)} and \textit{free(u)}.
There are no sink vertices for \textit{null-deref}.
According to Definitions~\ref{xdef:sl}--\ref{xdef:out},
we generate the following function summaries:
\begin{table}[h]
	\centering
	\begin{tabular}{llcl}
		\toprule
		\textbf{Function}      & \textbf{Summary Path} & \textbf{Label} & \textbf{Type} \\\midrule
		xmalloc                & ($p$, \textit{ret p})       & $0b111$        & transfer   \\\midrule
		\multirow{2}{*}{xfree} & ($u$, \textit{ret u})       & $0b111$        & input      \\
		                       & ($u$, \textit{free}($u$))   & $0b110$        & output     \\
	    \bottomrule
	\end{tabular}
\end{table}

The summary ($p$, \textit{ret p}) is labeled with $0b111$
because all three properties regard $p$ as the source.
The summary ($u$, \textit{ret u}) is also labeled with $0b111$
because the path does not contain any property-specific vertices and, thus, may be used for all three properties.
The summary ($u$, \textit{free}($u$)) is only labeled with $0b110$
because we do not regard \textit{free}($u$) as the sink for \textit{null-deref}.

When analyzing the main function,
we concatenate its intra-procedural paths with summaries from the callee functions
so as to generate a complete path.
For example, a concatenation is illustrated as below and its result is labeled by $0b110$,
meaning that the resulting path only works for \textit{double-free} and \textit{mem-leak}.
{
\begin{alignat*}{3}
	(p, \textit{ret } p)^{0b111} \circ (a) \circ (u, \textit{free}(u))^{0b110}& ~ && ~
	\\
	~ &\hspace{-3.8cm} =\enskip&& \hspace{-3.4cm} (p, \textit{ret } p, a, u, \textit{free}(u))^{0b111\&0b110}
	\\
	~ &\hspace{-3.8cm} =\enskip&& \hspace{-3.4cm} (p, \textit{ret } p, a, u, \textit{free}(u))^{0b110}
\end{alignat*}
}

We observe that using value-flow paths as function summaries has a significant advantage for checking multiple properties.
That is,
since value flow is a kind of fundamental program relations, it can be reused across different properties.
This is different from existing approaches that utilize state machine to model properties
and generate state-specific function summaries~\cite{fan2019smoke,das2002esp}. 
Since different properties usually have different states, compared to our value-flow-based function summaries,
such state-specific function summaries have fewer opportunities to be reused across properties.
\section{Implementation}
\label{sec:impl}

\begin{table*}[t]\small
	\centering
	\caption{Properties to Check in \toolname}
	\label{tab:checkers}
	\begin{tabulary}{1\textwidth}{clJ}
		\toprule
		\textbf{ID} & \textbf{Property Name}                        & \textbf{Brief Description}
		\\ 
		\midrule
		1  & \textsf{core.CallAndMessage}                      & Check for uninitialized arguments and null function pointers 
		\\ 
		2  & \textsf{core.DivideByZero}                        & Check for division by zero
		\\ 
		3  & \textsf{core.NonNullParamChecker}                 & Check for null passed to function parameters marked with nonnull
		\\ 
		4  & \textsf{core.NullDereference}                     & Check for null pointer dereference 
		\\ 
		5  & \textsf{core.StackAddressEscape}                  & Check that addresses of stack memory do not escape the function
		\\ 
		6  & \textsf{core.UndefinedBinaryOperatorResult}       & Check for the undefined results of binary operations
		\\ 
		7  & \textsf{core.VLASize (Variable-Length Array)}     & Check for declaration of VLA of undefined or zero size
		\\ 
		8  & \textsf{core.uninitialized.ArraySubscript}        & Check for uninitialized values used as array subscripts
		\\ 
		9  & \textsf{core.uninitialized.Assign}                & Check for assigning uninitialized values
		\\ 
		10 & \textsf{core.uninitialized.Branch}                & Check for uninitialized values used as branch conditions
		\\ 
		11 & \textsf{core.uninitialized.CapturedBlockVariable} & Check for blocks that capture uninitialized values
		\\ 
		12 & \textsf{core.uninitialized.UndefReturn}           & Check for uninitialized values being returned to callers
		\\ 
		13 & \textsf{cplusplus.NewDelete}                      & Check for C++ use-after-free
		\\ 
		14 & \textsf{cplusplus.NewDeleteLeaks}                 & Check for C++ memory leaks
		\\ 
		15 & \textsf{unix.Malloc}                              & Check for C memory leaks, double-free, and use-after-free
		\\ 
		16 & \textsf{unix.MismatchedDeallocator}               & Check for mismatched deallocators, e.g., new and free()
		\\ 
		17 & \textsf{unix.cstring.NullArg}                     & Check for null pointers being passed to C string functions like strlen
		\\ 
		18 & \textsf{alpha.core.CallAndMessageUnInitRefArg}    & Check for uninitialized function arguments
		\\ 
		19 & \textsf{alpha.unix.SimpleStream}                  & Check for misuses of C stream APIs, e.g., an opened file is not closed
		\\ 
		20 & \textsf{alpha.unix.Stream}                        & Check stream handling functions, e.g., using a null file handle in fseek
		\\ 
		\bottomrule
	\end{tabulary}
\end{table*}

In this section, we present the implementation details
as well as the properties to check in our framework.

\defpar{Path-sensitivity.} 
We have implemented our approach as a prototype tool called \toolname\ on top of \pin~\cite{shi2018pinpoint}.
Given the source code of a program, we first compile it to LLVM bitcode,\footnote{LLVM: \url{https://llvm.org/}}
on which our analysis is performed.
To achieve path-sensitivity, we build a path-sensitive value-flow graph and compute path conditions following the same method of \pin.
The path conditions in our analysis are first-order logic formulas over bit vectors. 
A program variable is modeled as a bit vector, of which the length is the bit width (e.g., 32) of the variable's type (e.g., int).
The path conditions are solved by Z3~\cite{de2008z3}, a state-of-the-art SMT solver, to determine path feasibility.

\defpar{Properties to check.} 
\toolname\ currently supports to check twenty C/C++ properties defined in \clang,
which are briefly introduced in Table~\ref{tab:checkers}.\footnote{More details of the properties can be found on \url{https://clang-analyzer.llvm.org/}.}
The twenty properties include all \clang's default C/C++ value-flow properties.
All other default C/C++ properties in \clang\ but not in \toolname\ are simple ones that do not require a path-sensitive analysis. 
For example, the property \textsf{security.insecureAPI.bcopy} requires \clang\ report 
a warning whenever a program statement calling the function \textit{bcopy()} is found.

\defpar{Parallelization.} 
Our analysis is performed in a bottom-up manner, in which a callee function is always analyzed before its callers.
Bottom-up compositional analysis is easy to run in parallel~\cite{xie2005scalable}.
Our special design for checking multiple properties does not prevent our analysis from parallelization.
As is common practice, 
in \toolname, functions that do not have calling relations are analyzed in parallel.

\defpar{Soundness.} 
We implement \toolname\ in a soundy manner~\cite{livshits2015defense}.
This means that
the implementation soundly handles most language features and, meanwhile, includes some well-known unsound design decisions as previous works~\cite{xie2005scalable,cherem2007practical,babic2008calysto,sui2014detecting,shi2018pinpoint}.
For example, in our implementation,
virtual functions are resolved by classic class hierarchy analysis~\cite{dean1995optimization}.
However,
we do not handle C style function pointers, inline assembly, and library functions.
We also follow the common practice to assume distinct function parameters do not alias each other~\cite{livshits2003tracking} and unroll each cycle twice on the call graph and the control flow graph.
These unsound choices significantly improve the scalability but have limited negative impacts on the bug-finding capability.

\section{Evaluation}
\label{sec:eval}

This section presents the systematic evaluation that demonstrates the high scalability, precision, and recall of our approach.


\begin{table}[t]
	\small
	\caption{Subjects for Evaluation}
	\label{tab:benchmarks}
	\begin{tabular}{llc|llc}
		\toprule
		\textbf{ID} & \textbf{Program} & \textbf{Size (KLoC)} & \textbf{ID}    & \textbf{Program}   & \textbf{Size (KLoC)} \\ 
		\midrule
		
		1   & mcf          & 2           & 13   & shadowsocks     & 32           \\ 
		2   & bzip2        & 3           & 14   & webassembly     & 75           \\ 
		3   & gzip         & 6           & 15   & transmission    & 88           \\
		4   & parser       & 8           & 16   & redis           & 101          \\ 
		5   & vpr          & 11          & 17   & imagemagick     & 358          \\ 
		6   & crafty       & 13          & 18   & python          & 434          \\ 
		7   & twolf        & 18          & 19   & glusterfs       & 481          \\ 
		8   & eon          & 22          & 20   & icu             & 537          \\ 
		9   & gap          & 36          & 21   & openssl         & 791          \\ 
		10  & vortex       & 49          & 22   & mysql           & 2,030        \\ 
		11  & perlbmk      & 73          &      &                 &              \\
		12  & gcc          & 135         & \multicolumn{2}{l}{\textbf{Total}} & 5,303                \\ 
		\bottomrule
	\end{tabular}
\end{table}

\subsection{Experimental Setup}

To demonstrate the scalability of our approach,
we compared the time and memory cost of \toolname\ with a series of existing industrial-strength static analyzers.
We also investigated their capability of finding real bugs, which confirms that our promising scalability is not achieved by sacrificing its bug-finding capability.

\defpar{Baseline approaches.}
First of all, we compared \toolname\ with \pin~\cite{shi2018pinpoint}, an open-source version of the most recent static analyzer of the same type.
Both of the two techniques are demand-driven, compositional, and sparse static analysis with the precision of path-sensitivity.
The difference is that \toolname\ exploits mutual synergies among different properties to speed up the analysis while \pin\ does not. 
In addition, we also conducted comparison experiments on the tools using abductive inference (\fbinfer) and
symbolic execution (\clang), both of which are open source and widely-used in industry.
This comparison aims to show that \toolname\ is competitive, 
as it consumes similar time and memory cost with \clang\ and \fbinfer, but is much more precise.
In the experiments, all tools were run with fifteen threads to take advantage of parallelization.

We also tried to compare with other static bug detection tools such as
\saturn~\cite{xie2005scalable}, \calysto~\cite{babic2008calysto}, \textsf{Semmle}~\cite{avgustinov2016ql}, \fortify, and \textsf{Klocwork}.\footnote{Klocwork: \url{https://www.roguewave.com/products-services/klocwork/}}
However, they are either unavailable or not runnable on the experimental environment we are able to set up.
The open-source static analyzer, \textsf{FindBugs},\footnote{Findbugs Static Analyzer: \url{http://findbugs.sourceforge.net/}} is not included in our experiments because it only works for Java while we focus on the analysis of C/C++ programs.
We do not compare with \tricoder~\cite{sadowski2015tricorder}, the static analysis platform from Google.
This is because the only C/C++ analyzer in it is \clang, which has been included in our experiments.

\defpar{Subjects for evaluation.}
To avoid possible biases on the benchmark programs, we include the standard and widely-used benchmark, SPEC CINT 2000\footnote{SPEC CPU2000: \url{https://www.spec.org/cpu2000/}} (ID = 1 $\sim$ 12 in Table \ref{tab:benchmarks}), in our evaluation.
At the same time,
in order to demonstrate the efficiency and effectiveness of \toolname\ on real-world projects,
we also include ten industrial-sized open-source C/C++ projects (ID = 13 $\sim$ 22 in Table \ref{tab:benchmarks}),
of which the size ranges from a few thousand to two million lines of code.


\defpar{Environment.}
All experiments were performed on a server with two Intel\textsuperscript{\copyright} Xeon\textsuperscript{\copyright} CPU E5-2698 v4 @ 2.20GHz (each has 20 cores) and 256GB RAM running Ubuntu-16.04.

\begin{figure*}[t]
	\centering
	\includegraphics[width=\textwidth]{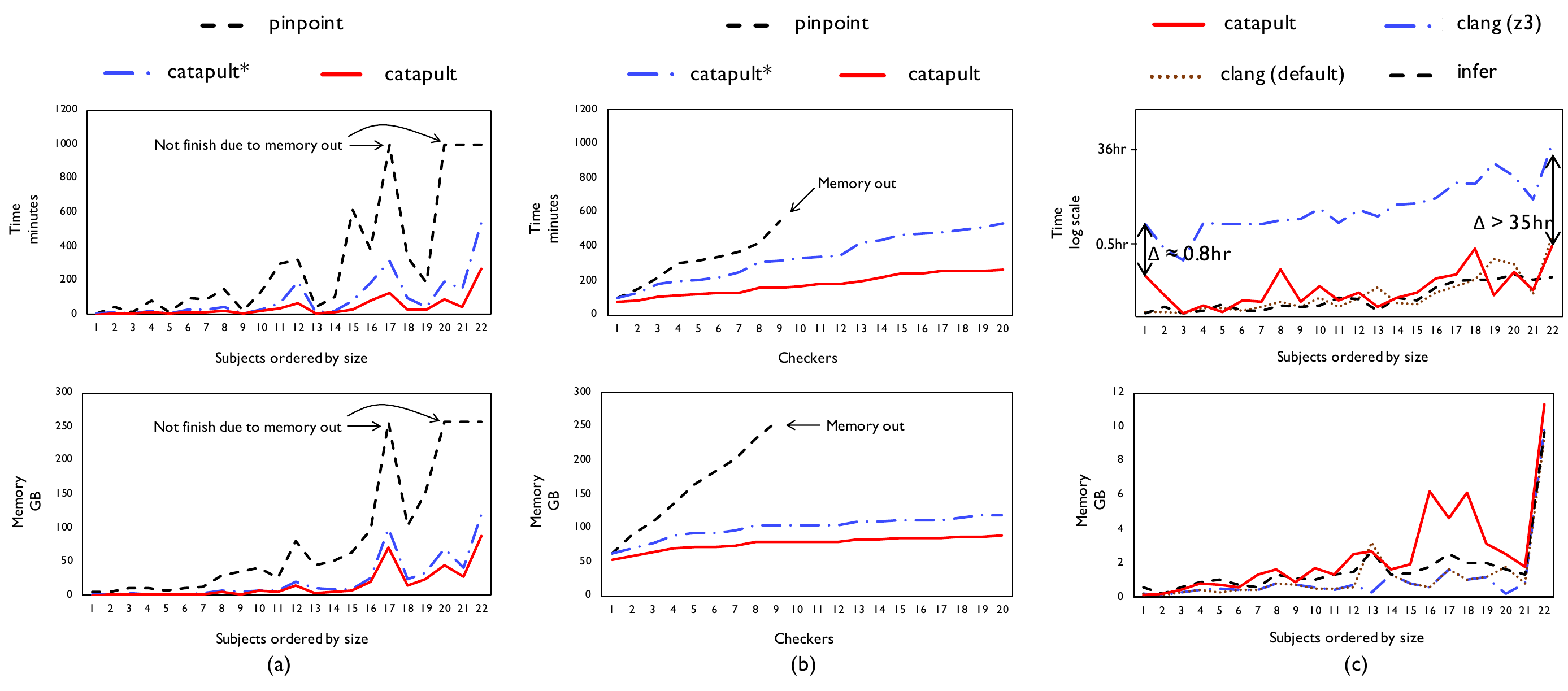}
	\caption{(a) Comparing time and memory cost with \pin. (b) The growth curves of the time and the memory overhead when comparing to \pin. (c) Comparing time and memory cost with \clang\ and \fbinfer.}
	\label{fig:eval_all}
\end{figure*}

\begin{table*}[t]
	\footnotesize
	\centering
	\caption{Effectiveness (\toolname\ vs. \pin, \clang, and \fbinfer)}
	\label{tab:bugs}
	\begin{tabular}{l?cc?cc}
		\toprule
		\multirow{2}{*}{\textbf{Program}}   & \multicolumn{2}{c?}{\textbf{\toolname}} & \multicolumn{2}{c}{\textbf{\pin}}   \\ \cline{2-5} 
		
		&    \textbf{\# Rep}  & \textbf{\# FP}  & \textbf{\# Rep} & \textbf{\# FP} \\ 
		
		\midrule
		shadowsocks                            & 9    &0    &9    &0  \\ 
		webassembly                            & 10   &2    &10   &2 \\ 
		transmission                           & 24   &2    &24   &2 \\ 
		redis                                  & 39   &5    &39   &5  \\ 
		imagemagick                            & 26   &8    &-    &-  \\ 
		python                                 & 48   &7    &48   &7  \\ 
		glusterfs                              & 59   &22   &59   &22 \\ 
		icu                                    & 161  &31   &-    &- \\ 
		openssl                                & 48   &15   &-    &-\\ 
		mysql                                  & 245  &88   &-    &- \\ \midrule
		\textbf{\% FP}                         & \multicolumn{2}{c?}{{26.9\%}}  & \multicolumn{2}{c}{{20.1\%}}    \\ 
		
		\bottomrule
		\multicolumn{5}{l}{}\\
		\multicolumn{5}{l}{}
	\end{tabular}
	\hspace{1cm}
	\begin{tabular}{l?cc?cc?cc?cc}
		\toprule
		\multirow{2}{*}{\textbf{Program}}   & \multicolumn{2}{c?}{\textbf{\toolname}} & \multicolumn{2}{c?}{\textbf{\clang~(Z3)}} & \multicolumn{2}{c?}{\textbf{\clang~(Default)}} & \multicolumn{2}{c}{\textbf{\fbinfer}$^\dagger$}   \\ \cline{2-9} 
		
		&    \textbf{\# Rep}  & \textbf{\# FP}  & \textbf{\# Rep} & \textbf{\# FP} & \textbf{\# Rep} & \textbf{\# FP} & \textbf{\# Rep} & \textbf{\# FP} \\ 
		
		\midrule
		shadowsocks             &   8    &2  & 24    &22 &25    &23    &15    &13  \\ 
		webassembly             &   4    &0  & 1     &0  &6     &2     &12    &12  \\ 
		transmission            &   31   &10 & 17    &12 &26    &21    &167*  &82  \\ 
		redis                   &   19   &6  & 15    &7  &32    &20    &16    &7   \\ 
		imagemagick             &   24   &7  & 34    &21 &78    &61    &34    &18  \\ 
		python                  &   37   &7  & 62    &40 &149*  &77    &82    &63  \\ 
		glusterfs               &   28   &5  & 0     &0  &268*  &82    &-     & -  \\ 
		icu                     &   55   &11 & 94    &67 &206*  &69    &248*  &71  \\ 
		openssl                 &   39   &19 & 44    &26 &44    &26    &211*  &85  \\ 
		mysql                   &   59   &20 & 271*  &59 &1001* &79    &258*  &80  \\ 
		\midrule
		\textbf{\% FP}          & \multicolumn{2}{c?}{{28.6\%}} & \multicolumn{2}{c?}{{64.9\%}}  & \multicolumn{2}{c?}{{75.7\%}}& \multicolumn{2}{c}{{78.6\%}}  \\ 
		
		\bottomrule
		\multicolumn{9}{l}{* We inspected one hundred randomly-sampled bug reports.}\\
		\multicolumn{9}{l}{$\dagger$ We fail to run the tool on glusterfs.}
	\end{tabular}
\end{table*}

\subsection{Comparing with Static Analyzer of the Same Type}
\label{subsec:eval:pin}

We first compared \toolname\ with \pin, the most recent static analyzer of the same type.
To demonstrate the power of the graph traversal plan and the optimization plan separately,
we also configured our approach by disabling the optimization plan, which is denoted as \toolname$^*$.

In this experiment,
we performed the whole program analysis.
That is,
we linked all compilation units in a project into a single file so that the static analyzers can perform cross-file analysis.
Before the analysis,
both \pin\ and \toolname\ need to build the value-flow graph as the program intermediate representation.
Since \toolname\ is built on top of \pin, the pre-processing time and the size of value-flow graph are the same for both tools,
which are almost linear to the size of a program~\cite{shi2018pinpoint}.
Typically, for MySQL, a program with about two million lines of code, 
it takes twenty minutes to build a value-flow graph with seventy million nodes and ninety million edges.
We omit the details of these data because it is not the contribution of this paper.

\defpar{Efficiency.} 
The time and memory cost of checking each benchmark program is shown in Figure \ref{fig:eval_all}a.
Owing to the inter-property-awareness, 
\toolname\ is about 8$\times$ faster than \pin\ and takes only 1/7 memory usage on average.
Typically, \toolname\ can finish checking MySQL in 5 hours,
which is aligned with the industrial requirement of finishing an analysis in 5 to 10 hours~\cite{bessey2010few, mcpeak2013scalable}.

When the optimization plan is disabled, 
\toolname$^*$ is about 3.5$\times$ faster than \pin\ and takes 1/5 memory usage on average.
Compared to the result of \toolname, it implies that 
the graph traversal plan and the optimization plan contribute to 40\% and 60\% of the time cost reduction, respectively.
Meanwhile, they contribute to 70\% and 30\% of the memory cost reduction, respectively.
As a summary, the two plans contribute similar to the time cost reduction, and the graph traversal plan is more important for the memory cost reduction because it allows us to abundantly share analysis results across different properties and avoid duplicate data storage.

Using the largest subject, MySQL, as an example,
Figure \ref{fig:eval_all}b illustrates the growth curves of the time and the memory overhead when the properties in Table \ref{tab:checkers} are added into the core engine one by one.
As illustrated, in terms of both time and memory overhead, 
\toolname\ grows much slower than \pin\ and, thus, scales up quite gracefully.

It is noteworthy that, except for the feature of inter-property-awareness, \toolname\ follows the same method of \pin\ to build value-flow graph and perform path-sensitive analysis. 
Thus, they have similar performance to check a single property.
\toolname\ performs better than \pin\ only when multiple properties are checked together.

\defpar{Effectiveness.}
Since both \toolname\ and \pin\ are inter-proce-durally path-sensitive,
as shown in Table \ref{tab:bugs}-Left, they produce a similar number of bug reports (\# Rep) and false positives (\# FP) for all the real-world programs
except for the programs that \pin\ fails to analyze due to the out-of-memory exception.

\subsection{Comparing with Other Static Analyzers}
\label{subsec:eval:others}

To better understand the performance of \toolname\ in comparison to other types of property-unaware static analyzers,
we also ran \toolname\ against two prominent and mature static analyzers, \clang\ (based on symbolic execution) and \fbinfer\ (based on abductive inference).
Note that 
\fbinfer\ does not classify the properties to check as Table \ref{tab:checkers} but targets at a similar range of properties, such as null dereference, memory leak, etc.

In the evaluation, \clang\ was run with two different configurations:
one is its default configuration where a fast but imprecise range-based solver is employed to solve path constraints,
and the other uses Z3~\cite{de2008z3}, a full-featured SMT solver, to solve path constraints.
To ease the explanation, we denote \clang\ in the two configurations as \clang~(Default) and \clang~(Z3), respectively.
Since \clang\ separately analyzes each source file and \fbinfer\ only has limited capability of detecting cross-file bugs,
for a fair comparison, 
all tools in the experiments were configured to check source files separately, and the time limit for analyzing each file is set to 60 minutes.
Since a single source file is usually small, we did not encounter memory issues in the experiment but missed a lot of cross-file bugs as discussed later.
Also, 
since we build value-flow graphs separately for each file and do not need to track cross-file value flows,
the time cost of building value-flow graphs is almost negligible. 
Typically, for MySQL, it takes about five minutes to build value-flow graphs for all files.
This time cost is included in the results discussed below.

Note that we did not change other default configurations of \clang\ and \fbinfer. 
This is because the default configuration is usually the best in practice. 
Modifying their default configuration may introduce more biases.

\defpar{Efficiency (\toolname\ vs. \clang~(Z3)).}
When both \toolname\ and \clang\ employ Z3 to solve path constraints,
they have similar precision (i.e., full path-sensitivity) in theory.
However, as illustrated in Figure \ref{fig:eval_all}c, \toolname\ is much faster than \clang\ and consumes similar memory for all the subjects.
For example, for MySQL, it takes about 36 hours for \clang\ to finish the analysis
while \toolname\ takes only half an hour but consumes similar memory space.
On average, \toolname\ is 68$\times$ faster than \clang\ at the cost of only 2$\times$ more memory to generate and store summaries.
In spite of the 2$\times$ more memory, both of them can finish the analysis in 12GB memory, which is affordable using a common personal computer.

\defpar{Efficiency (\toolname\ vs. \clang~(Default) and \fbinfer).}
As illustrated in Figure \ref{fig:eval_all}c, compared to \fbinfer\ and the default version of \clang,
\toolname\ takes similar (sometimes, a little higher) time and memory cost to check the subject programs.
For instance, for MySQL, the largest subject program, all three tools finish the analysis in 40 minutes and consume about 10GB memory.
With similar efficiency, \toolname, as a fully path-sensitive analysis, is much more precise than the other two.
The lower precision of \clang\ and \fbinfer\ leads to many false positives as discussed below.


\defpar{Effectiveness.}
In addition to the efficiency, 
we also investigate the bug-finding capability of the tools.
Table \ref{tab:bugs}-Right presents the results.
Since we only perform file-level analysis in this experiment, the bugs reported by \toolname\ is much fewer than those in Table \ref{tab:bugs}-Left.
Since validating each report may take tens of minutes, one day, or even longer,
we could not afford the time to manually inspect all of them.
Thus, we randomly sampled a hundred reports for the projects that have more than one hundred reports. 
We can observe from the results that, on average, 
the false positive rate of \toolname\ is much lower than \clang\ and \fbinfer.
In terms of recall, \toolname\ reports more true positives, which cover all those reported by \clang\ and \fbinfer.
\clang\ and \fbinfer\ miss many bugs because they make some trade-offs in exchange for efficiency.
For example, \clang\ often stops its analysis on a path after it finds the first possible bug.

Together with the results on efficiency,
we can conclude that \toolname\ is much more scalable than \clang\ and \fbinfer\ because they have similar time and memory overhead but \toolname\ is much more precise and able to detect more bugs.

\subsection{Detected Real Bugs}
\label{subsec:eval:effectiveness}

We note that the real-world software used in our evaluation is frequently scanned by commercial tools such as \textsf{Coverity SAVE}\footnote{Coverity Scan: \url{https://scan.coverity.com/projects/}} and, thus, is expected to have very high quality.
Nevertheless, 
\toolname\ still can detect many deeply-hidden software bugs that existing static analyzers, such as \pin, \clang, and \fbinfer, cannot detect.

At the time of writing, \confirmedbugnum\ previously-unknown bugs have been confirmed and fixed by the software developers,
including \confirmednpd\ null pointer dereferences, \confirmeduaf\ use-after-free or double-free bugs, \confirmedrl\ resource leaks, and \confirmedsae\ stack-address-escape bug.
Four of them even have been assigned CVE IDs due to their security impact.
We have made an online list for all bugs assigned CVE IDs or fixed by their original developers.\footnote{Detected real bugs: \url{https://qingkaishi.github.io/catapult.html}}

\begin{figure}[t]
	\centering
	\includegraphics[width=0.9\columnwidth]{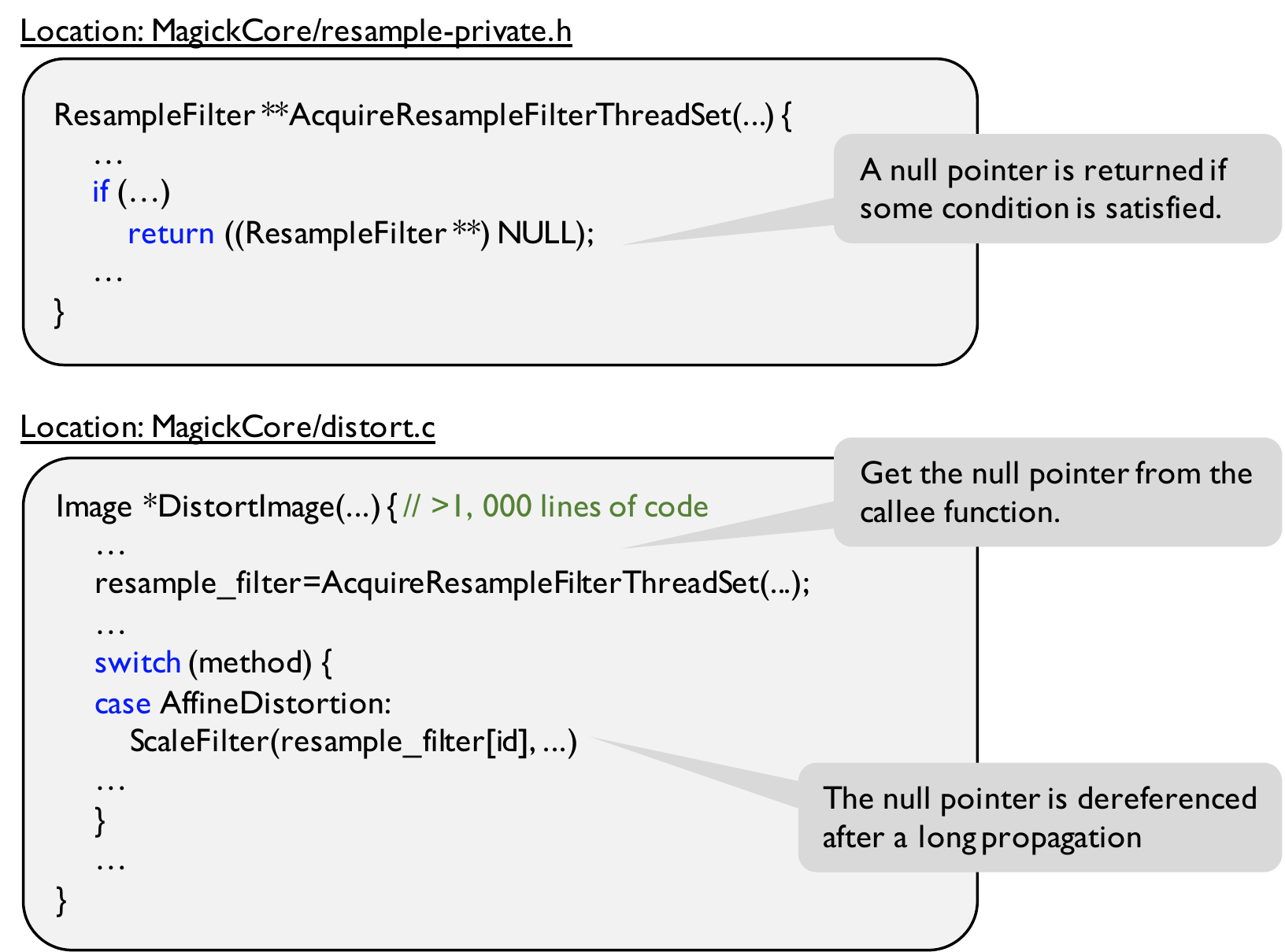}
	\caption{A null-dereference bug in ImageMagick.}
	\label{fig:realnpd}
\end{figure}

As an example, Figure \ref{fig:realnpd} presents a null-deference bug detected by \toolname\ in ImageMagick,
which is a software suite for processing images.
This bug is of high complexity, as it occurs in a function of more than 1,000 lines of code and the control flow involved in the bug spans across 56 functions over 9 files.

Since both \clang\ and \fbinfer\ make many unsound trade-offs to achieve scalability,
neither of them detects this bug.
\pin\ also cannot detect the bug because it is not memory-efficient and has to give up its analysis after the memory is exhausted.

\section{Related Work}
\label{sec:relatedwork}

To the best of our knowledge, a very limited number of existing static analyses have studied how to statically check multiple program properties at once,
although the problem is very important at an industrial setting.
\citet{goldberg2018efficient} make unsound assumptions and intentionally stop the analysis on a path after finding the first bug.
Apparently, the approach will miss many bugs, which violates our design goal.
Different from our approach that reduces unnecessary program exploration via cross-property optimization,
\citet{mordan2016checking} studied how to distribute computing resources,
so that the resources are not exhausted by a few properties.
\citet{cabodi2011optimized} studied the problem of checking multiple properties in the context of hardware model checking.
Their method has a similar spirit to our approach as it also tries to exploit mutual synergies among different properties.
However, it works in a different manner specially designed for hardware.
In order to avoid state-space explosion caused by large sets of properties, some other approaches studied how to decompose a set of properties into small groups~\cite{camurati2014split,apel2016fly}.
Owing to the decomposition, we cannot share the analysis results across different groups.
There are also some static analyzers such as \textsf{Semmle}~\cite{avgustinov2016ql} and \textsf{DOOP}~\cite{bravenboer2009strictly}
that take advantage of datalog engines for multi-query optimization.
However, they are usually not path-sensitive and
their optimization methods are closely related to the sophisticated datalog specifications.
In this paper, we focus on value-flow queries that can be simply specified as a quadruple and, thus, cannot benefit from the datalog engines.

\clang\ and \fbinfer\ currently are two of the most famous open-source static analyzers with industrial strength.
\clang\ is a symbolic-execution-based, exhaustive, and whole-program static analyzer. As a symbolic execution, it suffers from the path-explosion problem~\cite{king1976symbolic}.
To be scalable, it has to make unsound assumptions as in the aforementioned related work~\cite{goldberg2018efficient}, limit its capability of detecting cross-file bugs, and give up full path-sensitivity by default.
\fbinfer\ is an abstract-interpretation-based, exhaustive, and compositional static analyzer.
To be scalable, it also makes many trade-offs: giving up path-sensitivity and discarding sophisticated pointer analysis in most cases.
Similarly, \tricoder, the analyzer in Google, only works intra-procedurally in order to analyze large code base~\cite{sadowski2015tricorder, sadowski2018lessons}.

In the past decades, researchers have proposed many general techniques that can check different program properties but do not consider how to efficiently check them together~\cite{reps1995precise, ball2002slam, henzinger2002lazy, clarke2003behavioral,chaki2004modular,xie2005scalable, dillig2008sound,babic2008calysto, dillig2011precise, cho2013blitz, sui2016svf, shi2018pinpoint}. 
Thus, we study different problems.
In addition, there are also many techniques tailored only for a special program property, including null dereference~\cite{livshits2003tracking}, use after free~\cite{yan2018spatio},
memory leak~\cite{xie2005context,cherem2007practical, sui2014detecting, fan2019smoke}, buffer overflow~\cite{le2008marple}, etc.
Since we focus on the extensional scalability issue for multiple properties, our approach is different from them.

Value flow properties checked in our static analyzer are also related to well-known type-state properties~\cite{strom1983mechanisms,strom1986typestate}.
Generally, we can regard a value-flow property as a type-state property with at most two states.
Nevertheless,
value-flow properties have covered a wide range of program issues.
Thus, a scalable value-flow analyzer is really necessary and useful in practice.
Modeling a program issue as a value-flow property has many advantages.
For instance, \citet{cherem2007practical} pointed out that we can utilize the sparseness of value-flow graph to avoid tracking unnecessary value propagation in a control flow graph, thereby achieving better performance and outputting more concise issue reports.
In this paper,
we also demonstrate that using the value-flow-based model enables us to mitigate the extensional scalability issue.

\section{Conclusion}
\label{sec:conc}

We have presented \toolname, a scalable approach to checking multiple value-flow properties together.
The critical factor that makes our technique fast is to exploit the mutual synergies among the properties to check.
Since the number of program properties to check is quickly increasing nowadays, 
we believe that it will be an important research direction to study
how to scale up static program analysis for simultaneously checking multiple properties.

\begin{acks}
	The authors would like to thank the anonymous reviewers and Dr. Yepang Liu
	for their insightful comments.
	This work is partially funded by
	Hong Kong GRF16214515, GRF16230716, GRF16206517, and ITS/215/16FP grants.
	Rongxin Wu is the corresponding author.
\end{acks}

\balance
\bibliographystyle{ACM-Reference-Format}
\bibliography{bib/sigproc}

\onecolumn
\appendix

\newcommand{\IP}{\Pi_{\textsf{\textup{IP}}}}
\newcommand{\SL}{\Pi_{\textsf{\textup{SL}}}}
\newcommand{\TG}{\Pi}
\newcommand{\OUT}{\Pi_{\textsf{\textup{OUT}}}}
\newcommand{\IN}{\Pi_{\textsf{\textup{IN}}}}

\newcommand{\fpset}{V_{\textsf{\textup{fp}}}}
\newcommand{\apset}{V_{\textsf{\textup{ap}}}}
\newcommand{\frset}{V_{\textsf{\textup{fr}}}}
\newcommand{\arset}{V_{\textsf{\textup{ar}}}}
\newcommand{\srcset}{V_{\textsf{\textup{src}}}}
\newcommand{\sinkset}{V_{\textsf{\textup{sink}}}}

\newcommand{\fpelmt}{v_{\textsf{fp}}}
\newcommand{\apelmt}{v_{\textsf{ap}}}
\newcommand{\frelmt}{v_{\textsf{fr}}}
\newcommand{\arelmt}{v_{\textsf{ar}}}
\newcommand{\srcelmt}{v_{\textsf{src}}}
\newcommand{\sinkelmt}{v_{\textsf{sink}}}

\section{A Context-Free Grammar Model}

With no loss of generality, we assume the code in each function is in SSA form, where every variable has only one definition~\cite{cytron1989efficient}.
In a program, as existing work~\cite{li2011boosting,sui2014detecting,shi2018pinpoint},  
we say the value of a variable $a$ flows to a variable $b$ (or $b$ is data-dependent on $a$) if 
$a$ is assigned to $b$ directly (via an assignment, such as \textit{b=a}) or indirectly (via pointer dereferences, such as \textit{*p=a; q=p; b=*q}).
Then, the value-flow graph of a program is defined as below.

\begin{definition}[Value-Flow Graph]
	A value-flow graph is a directed graph $G = (V, E)$, where $V$ and $E$ are defined as following:
	\begin{itemize}
		\item $V$ is a set of vertices, each of which is denoted by $v@s$, meaning that the variable $v$ is defined or used in the statement $s$.
		
		\item $E\subseteq V\times V$ is a set of edges, each of which represents a data dependence relation or value flow. 
		$(v_1@s_1, v_2@s_2)\in E$ means that the value of $v_1@s_1$ flows to $v_2@s_2$.
	\end{itemize}
	\label{def:vfg}
\end{definition}

We say $\pi =   (v_0@s_0, v_1@{s_1}, \cdots, v_n@{s_n} )$ is a value-flow path
if and only if 
the sequence represents a path on the value-flow graph.
We use $\pi[i]$ to represent $v_i@{l_i}$ on the path if $0\le i\le n$. 
Specifically, we use $\pi[-1]$ to represent the last element of $\pi$.
A value-flow path $\pi_2$ can be concatenated to the other one $\pi_1$, denoted as $\pi_1\pi_2$,
if and only if $(\pi_1[-1], \pi_2[0]) \in E$.
Given $V_1, V_2, V_3, V_4\subseteq V$, we use $\Pi(V_1, V_2)$ to represent the set of value-flow paths from a vertex in $V_1$ to a vertex in $V_2$.
The concatenation of two value-flow paths then can be extended to two sets:
$\Pi(V_1, V_2)\Pi(V_3, V_4)=\{\pi_1\pi_2:  \pi_1\in \Pi(V_1, V_2) \land \pi_2\in \Pi(V_3, V_4) \land (\pi_1[-1], \pi_2[0])\in E\}.$

In the following definitions, we use $\fpset, \apset, \frset$, and $\arset$ to represent four special vertex subsets.
$\fpset$ and $\apset$ represent the sets of formal and actual parameters, respectively.
$\frset$ and $\arset$ represent the sets of formal and actual return values, respectively. We refer to the return value at a return statement as the formal return value (e.g., $\textit{v@return v}$) and
the return value at a call statement as the actual return value (e.g., $\textit{v@v=func()}$).
All proofs in this subsection are put in Appendix B.

\begin{definition}[Intra-Procedural Value-Flow Paths, $\IP(V_i, V_j)$]
	Given a value-flow graph $G=(V, E)$ and $V_i, V_j\subseteq V$,
	a value-flow path $\pi \in \IP(V_i, V_j)$ iff. $\pi \in \Pi(V_i, V_j)$ and
	$\forall m\ge 0, n\ge 0$: $\pi[m]$ and $\pi[n]$ are in the same function.\footnote{
		For simplicity, when we say $v_1@s_1$ and $v_2@s_2$ are in the same function,
		we mean that they are in the same function under the same calling context.
	}
	\label{def:ip}
\end{definition}


As defined below, a same-level value-flow path starts and ends in the same function, but may go through some callee functions.


\begin{definition}[Same-Level Value-Flow Paths, $\SL(V_i, V_j)$]
	\label{def:sl}
	Given a value-flow graph $G=(V, E)$ and $V_i, V_j\subseteq V$,
	a value-flow path $\pi \in \SL(V_i, V_j)$ iff. $\pi \in \Pi(V_i, V_j)$ and
	$\pi[0]$ is in the same function with $\pi[-1]$.
\end{definition}

\begin{figure}
	\centering
	\includegraphics[width=\columnwidth]{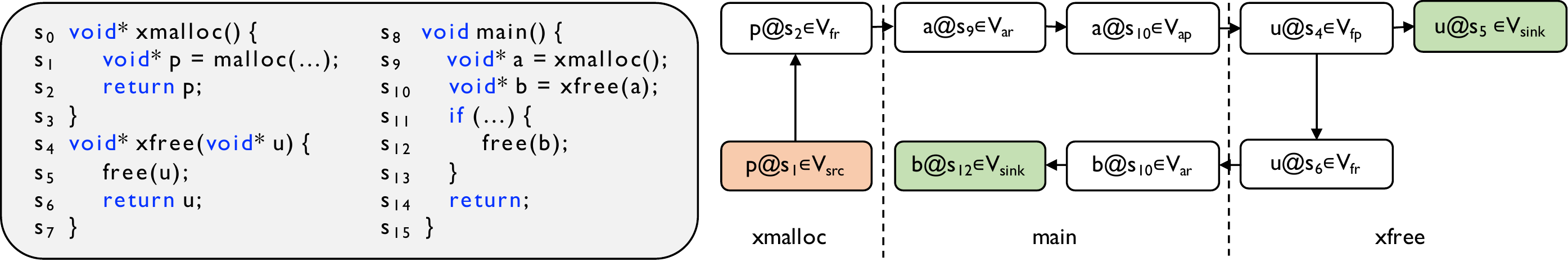}
	\caption{Code and its value-flow graph for explaining Examples \ref{ex:sl} - \ref{ex:in}}
	\label{fig:app_example}
\end{figure}

\begin{example}[Same-Level Value-Flow Paths, $\SL(V_i, V_j)$]
	The value-flow path $(a@s_{10}, u@s_4,$ $u@s_6, b@s_{10})$ in Figure \ref{fig:app_example}
	is a same-level value-flow path because the head of the path $a@s_{10}$ is in the same function with the tail $b@s_{10}$.
	\label{ex:sl}
\end{example}


\begin{lemma}[Same-Level Value-Flow Paths, $\SL(V_i, V_j)$]
	$\SL(V_i, V_j)$ can be generated using the following productions:
	\begin{eqnarray}
	\SL(V_i, V_j) &\rightarrow& \IP( V_i, V_j ) 
	\label{eqn:sl1}\\
	\SL(V_i, V_j) &\rightarrow& \IP (V_i, \apset ) \SL(\fpset, \frset) \SL (\arset, V_j )
	\label{eqn:sl2}
	\end{eqnarray}
	\label{lemma:samelevel}
\end{lemma}


An output value-flow path, which is defined below, indicates that a checker-specific source escapes to its caller functions or upper-level caller functions.


\begin{definition}[Output Value-Flow Paths, $\OUT(\srcset, \frset)$]
	\label{def:out}
	Given a value-flow graph $G=(V, E)$,
	a value-flow path $\pi \in \OUT(\srcset, \frset)$ iff. $\pi \in \Pi(\srcset, \frset)$
	and $\pi[-1]$ is in the same function with $\pi[0]$ or in the (upper-level) callers of $\pi[0]$'s function.
\end{definition}


\begin{example}[Output Value-Flow Paths, $\OUT(\srcset, \frset)$]
	In Figure \ref{fig:app_example},
	$(p@s_1, p@s_2)$ is an output value-flow path, 
	because the source vertex $p@s_1$ flows to $p@s_2$, which is a formal return value and in the same function with $p@s_1$.
	\label{ex:out}
\end{example}

%
\begin{lemma}[Output Value-Flow Paths, $\OUT(\srcset, \frset)$]
	$\OUT(\srcset, \frset)$
	can be generated using the following productions:
	\begin{eqnarray}
	\OUT(\srcset, \frset) &\rightarrow& \SL(\srcset, \frset) 
	\label{eqn:out1}\\
	\OUT(\srcset, \frset) &\rightarrow& \OUT(\srcset, \frset)\SL(\arset, \frset)
	\label{eqn:out2}
	\end{eqnarray}
	\label{lemma:output}
\end{lemma}


An input value-flow path, as defined below, indicates that a formal parameter of a function $f$ may flow to a sink vertex in $f$ or $f$'s callees.
A source vertex in $f$'s caller functions may propagate to the sink through the formal parameter.


\begin{definition}[Input Value-Flow Paths, $\IN(\fpset, \sinkset)$]
	\label{def:in}
	Given a value-flow graph $G=(V, E)$,
	a value-flow path $\pi \in \IN(\fpset, \sinkset)$ iff. $\pi \in \Pi(\fpset, \sinkset)$
	and $\pi[-1]$ is in the same function with $\pi[0]$ or in the (lower-level) callees
	of $\pi[0]$'s function.
\end{definition}


\begin{example}[Input Value-Flow Paths, $\IN(\fpset, \sinkset)$]
	In Figure \ref{fig:app_example}, 
	$(u@s_4, u@s_5)$ is an input value-flow path because it starts with a formal parameter $u@s_4$
	and ends with a sink vertex $u@s_5$ in the same function.
	\label{ex:in}
\end{example}


\begin{lemma}[Input Value-Flow Paths, $\IN(\fpset, \sinkset)$]
	$\IN(\fpset, \sinkset)$
	can be generated using the following productions:
	\begin{eqnarray}
	\IN(\fpset, \sinkset) &\rightarrow& \SL(\fpset, \sinkset)
	\label{eqn:in1}\\
	\IN(\fpset, \sinkset) &\rightarrow& \SL(\fpset, \apset)\IN(\fpset, \sinkset)
	\label{eqn:in2}
	\end{eqnarray}
	\label{lemma:input}
\end{lemma}


\begin{lemma}[Target Value-Flow Paths, $\TG(\srcset, \sinkset)$]
	Given a checker $(\srcset, \sinkset, \mathcal{F}_p, \mathcal{A}_q)$, 
	the set of target value-flow paths $\TG(\srcset, \sinkset)$ can be generated using the following productions:
	\begin{alignat}{3}
		\TG(\srcset, \sinkset) \enskip&\rightarrow\enskip&& \SL(\srcset, \sinkset)
		\label{eqn:tg1}\\
		\TG(\srcset, \sinkset) \enskip&\rightarrow\enskip&& \SL(\srcset, \apset) \IN(\fpset, \sinkset)
		\label{eqn:tg2}\\
		\TG(\srcset, \sinkset) \enskip&\rightarrow\enskip&& \OUT(\srcset, \frset) \SL(\arset, \sinkset)
		\label{eqn:tg3}\\
		\TG(\srcset, \sinkset) \enskip&\rightarrow\enskip&& \OUT(\srcset, \frset) \SL(\arset, \apset) \IN(\fpset, \sinkset)
		\label{eqn:tg4}
	\end{alignat}
	
	\label{lemma:target}
\end{lemma}

The context-free grammar (Productions (\ref{eqn:sl1}) - (\ref{eqn:tg4})) implies that there is a Turing machine (or an algorithm) that can generate the target set $\TG(\srcset, \sinkset)$ of value-flow paths by concatenating various value-flow paths~\cite{hopcroft2006automata}, which sets the foundation for our compositional analysis.
According to the grammar, we can prove that, in the compositional analysis, it is sufficient to generate three kinds of function summaries, i.e., value-flow paths in $\SL(\fpset, \frset)$, $\IN(\fpset, \sinkset)$, and $\OUT(\srcset, \frset)$.
The sufficiency is described as the following theorem and proved in the appendices.

\begin{theorem}[Summary Sufficiency]
	Any target value-flow path in $\Pi(\srcset, \sinkset)$ can be written as the concatenation of (1) a function's intra-procedural value-flow paths, and (2) value-flow paths in $\SL(\fpset, \frset)$, $\IN(\fpset, \sinkset)$, and $\OUT(\srcset, \frset)$ from its callees.
	\label{theorem:suff}
\end{theorem}

\section{Proofs}

Given a global value-flow graph $G=(V, E)$,
we now explain the proofs of the lemmas and theorems in the paper.
To ease the explanation, we use $\srcelmt, \sinkelmt, \fpelmt, \apelmt$, $\frelmt$ and $\arelmt$ to
represent elements in the sets $\srcset, \sinkset, \fpset, \apset$, $\frset$ and $\arset$, respectively.
Sometimes we add superscript to them, e.g., $\fpelmt^0, \fpelmt^1, \cdots$, to represent a list of such elements with indices.
In the proofs, when we say two elements in $V$ are in the same function, we mean they are in the same function as well as the same calling context.

\bigskip

Proof of Lemma \ref{lemma:samelevel}.
\begin{proof}
	~
	
	(1) \textbf{Prove}: $\SL(V_i, V_j) \supseteq \IP( V_i, V_j ) \cup \IP (V_i, \apset ) \SL(\fpset, \frset) \SL (\arset, V_j )$.
	
	First, according to Definitions \ref{def:ip} and \ref{def:sl}, it is straightforward to conclude $\forall \pi \in \IP(V_{i}, V_{j}) : \pi \in \SL(V_{i}, V_{j})$.
	
	Second, $\forall \pi \in \IP (V_i, \apset ) \SL(\fpset, \frset)\SL (\arset, V_j )$, $\pi$ can be written as $\pi_1\pi_2\pi_3$
	where $\pi_1\in \IP (V_i, \apset )$, $\pi_2\in \SL(\fpset, \frset)$, and $\pi_3\in \SL (\arset, V_j )$.
	Since $\pi_1[-1]$ is an actual parameter and $\pi_2[0]$ is a formal parameter, the concatenation of $\pi_1$ and $\pi_2$
	means that we enter into a callee function, say \textit{foo}.
	Since the formal parameter $\pi_2[0]$ and the formal return value $\pi_2[-1]$ are in the same function, 
	the concatenation of $\pi_2$ and $\pi_3$ means that we exit from the callee \textit{foo}.
	Thus, $\pi_1[-1]$ and $\pi_3[0]$ are actually the actual parameter and actual return value at the same call site.
	Since $\pi_3$ is a same-level path, $\pi_3[0]$ and $\pi_3[-1]$ are in the same function.
	Hence, $\pi_1[0]$ and $\pi_3[-1]$ is in the same function, meaning that $\pi = \pi_1\pi_2\pi_3\in \SL(V_i, V_j)$.	
	
	\bigskip
	
	(2) \textbf{Prove}: $\SL(V_i, V_j) \subseteq \IP( V_i, V_j ) \cup \IP (V_i, \apset ) \SL(\fpset, \frset) \SL (\arset, V_j )$.
	
	$\forall \pi \in \SL(V_i, V_j)$, $\pi$ can be intra-procedural or inter-procedural.
	If it is an intra-procedural path, then $\pi\in \IP( V_i, V_j )$.
	
	For an inter-procedural path $\pi \in \SL(V_i, V_j)$,
	since $\pi[0]$ and $\pi[-1]$ are in the same function, 
	the value of $\pi[0]$ must flow to the other function, say \textit{foo}, and, then, flow back.
	The function \textit{foo} must be callee function because because if a value is returned to the caller function, it cannot flow back to the same function in the same calling context.
	Therefore, a path in $\SL(V_i, V_j)$ must be in the following form, where $v_i\in V_i, v_j\in V_j$, the value flow $(\apelmt^{p}, \fpelmt^{p+1})$ is a function call,
	and the value flow $(\frelmt^{p+1}, \arelmt^{p+2})$ is a function return, $(p=1,2,\cdots, 2n+1)$.
	$$
	\underbrace{
		( v_i... \apelmt^1 )
	}_{\IP( V_i, \apset) }
	\underbrace{
		( \fpelmt^2... \frelmt^2 )
	}_{\SL(\fpset, \frset)}
\underbrace{
		\underbrace{
			( \arelmt^3... \apelmt^3 )
		}_{\IP (\arset, \apset )}
		\underbrace{
			( \fpelmt^4... \frelmt^4 )
		}_{\SL(\fpset, \frset)}
		\underbrace{
			( \arelmt^5... \apelmt^5 )
		}_{\IP (\arset, \apset )}
		\underbrace{
			( \fpelmt^6... \frelmt^6 )
		}_{\SL(\fpset, \frset)}
		\cdots
		\underbrace{
			( \arelmt^{2n+1}... \apelmt^{2n+1} )
		}_{\IP (\arset, \apset )}
		\underbrace{
			( \fpelmt^{2n+2}... \frelmt^{2n+2} )
		}_{\SL(\fpset, \frset)}
	\underbrace{
		( \arelmt^{2n+3}... v_j )
	}_{\IP( \arset, V_j)}	
}_{\SL (\arset, V_j )}
	$$
	Thus, if $\pi\in \SL(V_i, V_j)$ is inter-procedural, $\pi\in \IP (V_i, \apset ) \SL(\fpset, \frset) \SL (\arset, V_j )$.
\end{proof}

Proof of Lemma \ref{lemma:output}.
\begin{proof}
	~
	
	(1) \textbf{Prove}: $\OUT(\srcset, \frset) \supseteq \SL(\srcset, \frset) \cup \OUT(\srcset, \frset)\SL(\arset, \frset)$
	
	First, according to Definitions \ref{def:sl} and \ref{def:out}, it is straightforward to conclude $\forall \pi \in \SL(\srcset, \frset) : \pi \in \OUT(\srcset, \frset)$.
	
	Second, $\forall \pi \in \OUT(\srcset, \frset)\SL(\arset, \frset)$, it can be written as $\pi_1\pi_2$ where $\pi_1\in \OUT(\srcset, \frset)$ and $\pi_2\in \SL(\arset, \frset)$.
	Since $\pi_1[-1]$ is a formal return value, the concatenation of $\pi_1$ and $\pi_2$ means a function return. Thus, $\pi_2$ is in the caller function of the source value $\pi_1[0]$.
	According to Definition~\ref{def:out}, $\pi = \pi_1\pi_2\in \OUT(\srcset, \frset)$.
	
	\bigskip
	
	(2) \textbf{Prove}: $\OUT(\srcset, \frset) \subseteq \SL(\srcset, \frset) \cup \OUT(\srcset, \frset)\SL(\arset, \frset)$
	
	According to Definition \ref{def:out}, $\forall \pi\in \OUT(\srcset, \frset)$, it must be in the form $(\srcelmt, ..., \frelmt)$.
	And $\frelmt$ is in the same function with $\srcelmt$ or $\frelmt$ is in certain (direct or indirect) callers of $\srcelmt$.
	If $\frelmt$ is in the same function with $\srcelmt$, it means $\pi$ is a same-level path and, thus, is in the set $\SL(\srcset, \frset)$.
	
	If $\frelmt$ is in a caller function of $\srcelmt$,
	since a program state only can transit to its callers via function return, $(\srcelmt,...,\frelmt)$ can be split into multiple parts concatenated by return value flows $(\frelmt, \arelmt)$:
	$$
	\underbrace{
		\underbrace{
			( \srcelmt...\frelmt^0)
		}_{\SL(\srcset, \frset)}
		\underbrace{
			( \arelmt^1...\frelmt^1)
		}_{\SL(\arset, \frset)}
		\underbrace{
			( \arelmt^2...\frelmt^2)
		}_{\SL(\arset, \frset)}
		\cdots
		\underbrace{
			( \arelmt^n...\frelmt^n)
		}_{\SL(\arset, \frset)}
	}_{\OUT(\srcset, \frset)}
	\underbrace{( \arelmt^{n+1}...\frelmt)}_{\SL(\arset, \frset)}
	$$
	Thus, $\forall \pi = (\srcelmt, ..., \frelmt) \in \OUT(\srcset, \frset)$, if $\frelmt$ is in a caller function of $\srcelmt$,
	$\pi$ must be in the set $\OUT(\srcset, \frset)\SL(\arset, \frset)$.
\end{proof}

Proof of Lemma \ref{lemma:input}.
\begin{proof}
	~
	
	(1) \textbf{Prove}: $\IN(\fpset, \sinkset) \supseteq \SL(\fpset, \sinkset)\cup \SL(\fpset, \apset)\IN(\fpset, \sinkset)$
	
	First, according to Definitions \ref{def:sl} and \ref{def:in}, it is straightforward to conclude $\forall \pi \in \SL(\fpset, \sinkset) : \pi \in \IN(\fpset, \sinkset)$.
	
	Second, $\forall \pi \in \SL(\fpset, \apset)\IN(\fpset, \sinkset)$, it can be written as $\pi_1\pi_2$ where $\pi_1\in \SL(\fpset, \apset)$ and $\pi_2\in \IN(\fpset, \sinkset)$.
	Since $\pi_1[-1]$ is an actual parameter, the concatenation of $\pi_1$ and $\pi_2$ means a function call. Thus, $\pi_2$ is in the callee function of the source value $\pi_1[0]$.
	According to Definition~\ref{def:in}, $\pi = \pi_1\pi_2\in \IN(\fpset, \sinkset)$.
	
	\bigskip
	
	(2) \textbf{Prove}: $\IN(\fpset, \sinkset) \subseteq \SL(\fpset, \sinkset)\cup \SL(\fpset, \apset)\IN(\fpset, \sinkset)$
	
	According to Definition \ref{def:in}, $\forall \pi \in \IN(\fpset, \sinkset)$, $\pi$ must be in the form $(\fpelmt, ..., \sinkelmt)$,
	where (1)~$\sinkelmt$ is in the same function with $\fpelmt$ or (2) $\sinkelmt$ is in certain (direct or indirect) callee of $\fpelmt$.
	If $\sinkelmt$ is in the same function with $\fpelmt$, 
	$(\fpset ... \sinkset) \in \SL(\fpset, \sinkset)$.
	
	If $\sinkset$ is in some callee of $\fpset$, 
	since a state only can transit to its callees via function calls,
	$( \fpset...\sinkset)$ can be split into multiple parts concatenated by function-call value flows $(\apelmt, \fpelmt)$:
	$$
	\underbrace{( \fpelmt...\apelmt^0)}_{\SL(\fpset, \apset)}
	\underbrace{
		\underbrace{
			( \fpelmt^1...\apelmt^1)
		}_{\SL(\fpset, \apset)}
		\underbrace{
			( \fpelmt^2...\apelmt^2)
		}_{\SL(\fpset, \apset)}
		\cdots
		\underbrace{
			( \fpelmt^n...\apelmt^n)
		}_{\SL(\fpset, \apset)}
		\underbrace{
			( \fpelmt^{n+1}...\sinkelmt)
		}_{\SL(\fpset, \apset)}
	}_{\IN(\fpset, \sinkset)}
	$$
	Thus, $\forall \pi = (\srcelmt, ..., \frelmt) \in \IN(\fpset, \sinkset)$, if $\sinkset$ is in some callee of $\fpset$,
	$\pi$ must be in the set $\SL(\fpset, \apset)\IN(\fpset, \sinkset)$.
\end{proof}

Proof of Lemma \ref{lemma:target}.
\begin{proof}
	To ease explanation, let us use {\sf RHS} to represent the set of value-flow paths in the right-hand side of Productions (\ref{eqn:tg1})-(\ref{eqn:tg4}).
	
	(1) \textbf{Prove}: $\TG(\fpset, \sinkset) \supseteq \textsf{RHS}$
	
	According to the definitions, it is straightforward to conclude that $\forall \pi \in {\sf RHS}: \pi \in \TG(\srcset, \sinkset)$.
	
	\bigskip
	
	(2) \textbf{Prove}: $\TG(\fpset, \sinkset) \subseteq \textsf{RHS}$
	
	Given a target value-flow path $\pi=( \srcelmt, ..., \sinkelmt)$,
	we search the path from $v = \sinkelmt$ in reverse order until the last vertex $\pi[i]$ satisfying the searching condition:
	$\pi[i]$ and $v$ are in the same function and $\pi[i]\in \fpset$.
	If we find such a $\pi[i]$, set $\sigma = \pi[i-1]$, which must be in $\apset$, and continue the search. 
	Otherwise, we stop the search. 
	Accordingly, we can rewrite the target value-flow path as below, where the arrows indicate the searching process:
	$$
	( \srcelmt... \apelmt^n, 
	\underbrace{
		\overleftarrow{
			\underbrace{
				\fpelmt^n... \apelmt^{n-1}
			}_{\SL(\fpset, \apset)}
		}
		......
		\overleftarrow{
			\underbrace{
				\fpelmt^2... \apelmt^1
			}_{\SL(\fpset, \apset)}
		}, 
		\overleftarrow{
			\underbrace{
				\fpelmt^1... \apelmt^0
			}_{\SL(\fpset, \apset)}
		}, 
		\overleftarrow{
			\underbrace{
				\fpelmt^0... \sinkelmt)
			}_{\SL(\fpset, V_{\bullet sink})}
		}
	}_{\IN(\fpset, \sinkset)}
	$$
	
	\textbf{Case (1):} If $\srcelmt$ is in the same function with $\apelmt^n$,
	it can be written as below and, thus, we have Production (\ref{eqn:tg2}).
	$$
	\underbrace{
		( \srcelmt... \apelmt^n 
	}_{\SL(\srcset, \apset)},
	\underbrace{
		 \fpelmt^n...\sinkelmt )
	}_{\IN(\fpset, \sinkset)}
	$$
	
	\textbf{Case (2):} If $\srcelmt$ is not in the same function with $\apelmt^n$,
	$\srcelmt$ must be in a callee of $\apelmt^n$, otherwise, 
	it is contradicted with the assumption that we stop the search at the $\fpelmt^n$.
	Since $\srcelmt$ is in a callee of $\apelmt^n$, the target value-flow path can be written as below. Thus, we have Production (\ref{eqn:tg3}).
	$$
	 \underbrace{( \srcelmt... \frelmt}_{\OUT(\srcset, \frset)},\underbrace{\arelmt... \apelmt^n}_{\SL(\arset, \apset},
		\underbrace{
			\fpelmt^n...\sinkelmt )
		}_{\IN(\fpset, \sinkset)}
	$$
	
	\textbf{Case (3):}
	During the search, if we never find any $\pi[i]$ satisfying the searching condition 
	and $\srcelmt$ and $\sinkelmt$ are in the same function.
	According to Definition \ref{def:sl}, the target value-flow path is in $\SL(\srcset, \sinkset)$. 
	Thus, we have Production (\ref{eqn:tg1}).
	
	\textbf{Case (4):} During the search, if we never find any $\pi[i]$ satisfying the searching condition and $\srcelmt$ and $\sinkelmt$ are in different functions,
	similar to Case (2), $\srcelmt$ must be in some callee of $\sinkelmt$. 
	Thus, the target value-flow path can be rewritten as below and we get Production (\ref{eqn:tg4}).
	$$
	 \underbrace{( \srcelmt... \frelmt}_{\OUT(\srcset, \frset)},\underbrace{\arelmt... \sinkelmt)}_{\SL(\arset, \sinkset)}
	$$
\end{proof}

Proof of Theorem \ref{theorem:suff}.
\begin{proof}
A target value-flow path in $\TG(\srcset, \sinkset)$ generated by Production (\ref{eqn:tg1}) can be rewritten according to the following two different derivations, in which $*$ is the Kleene star operation and $1\le k \le n$.	
	\small
	\begin{eqnarray*}
		\TG(\srcset, \sinkset) &\rightarrow& \underline{\SL(\srcset, \apset)} \IN(\fpset, \sinkset) \\
		~&~&\qquad\underbrace{\pi_1}\qquad\enspace\underbrace{\pi_2}\\
		&\rightarrow &\IP(\srcset, \apset) \IN(\fpset, \sinkset)       \\
		~&~&~\\                      
		\TG(\srcset, \sinkset) &\rightarrow& \underline{\SL(\srcset, \apset)} \IN(\fpset, \sinkset)\\
		&\rightarrow& \IP (\srcset, \apset ) \SL(\fpset, \frset) \underline{\SL (\arset, \apset )} \IN(\fpset, \sinkset)\\
		&\rightarrow& \IP (\srcset, \apset ) \SL(\fpset, \frset) (\IP (\arset, \apset ) \SL(\fpset, \frset))* \underline{\SL (\arset, \apset )}  \IN(\fpset, \sinkset)\\
		~&~&\qquad\underbrace{\pi_1}\qquad\enspace\underbrace{\pi_2}\qquad\enspace\underbrace{\pi_{2k+1}}\qquad\;\underbrace{\pi_{2k+2}}\qquad\quad\enspace\underbrace{\pi_{2n+3}}\qquad\enspace\underbrace{\pi'}\\
		&\rightarrow& \IP (\srcset, \apset ) \SL(\fpset, \frset) (\IP (\arset, \apset ) \SL(\fpset, \frset))* \IP (\arset, \apset )  \IN(\fpset, \sinkset)
	\end{eqnarray*}
	\normalsize
	
	In the first derivation, Production (\ref{eqn:sl1}) is applied to the underline part and a target value-flow path $\pi$ then can be rewritten as $\pi_1\pi_2$.
	Here, $\pi_1$ is an intra-procedural path in a function and $\pi_2$ must come from one of the function's callees.
	This is because $\pi_1[-1]\in \apset$ and $\pi_2[0]\in \fpset$, which means that the value-flow $(\pi_1[-1], \pi_2[0])$ is a function call.
	
	In the second derivation, Production (\ref{eqn:sl2}) is applied multiple times to the underlined parts in the first two rows and, finally, Production (\ref{eqn:sl1}) is applied to the underlined part in the third row.
	Following this derivation, a  target value-flow path is split into more parts where any triple $\pi_{2k-1}\pi_{2k}\pi_{2k+1}$ means entering into a callee function ($\pi_{2k-1}[-1]\in \apset$ and $\pi_{2k}[0]\in \fpset$) and returning back ($\pi_{2k}[-1]\in \frset$ and $\pi_{2k+1}[0]\in \arset$).
	Thus, all intra-procedural parts (i.e., $\pi_1$ and all $\pi_{2k+1}$) are in the same function and the other parts (i.e., $\pi'$ and all $\pi_{2k}$) are from callees.
	
	Thus, based on Production (\ref{eqn:sl1}), a target value-flow path can be rewritten as the concatenation 
	of the intra-procedural paths of a function and other paths in $\IN(\fpset, \sinkset)$ and $\SL(\fpset, \frset)$ from callees.
	
	\bigskip
	
	In a similar manner, after analyzing Productions (\ref{eqn:tg2})-(\ref{eqn:tg4}),
	we can conclude that when analyzing a function, it is sufficient to have value-flow paths in $\IN(\fpset, \sinkset)$, $\OUT(\srcset, \frset)$ and $\SL(\fpset, \frset)$ from callees.
\end{proof}

\end{document}